\documentstyle[prl,multicol,epsf,aps]{revtex}
\draft

\title{Semiclassical Time Evolution of the Density Matrix\break
and Tunneling} 

\author{Joachim Ankerhold and Hermann Grabert}
\address{
Fakult{\"a}t f{\"u}r Physik, Albert-Ludwigs-Universit{\"a}t
Freiburg,
 Hermann-Herder-Stra{\ss}e 3, D-79104 Freiburg, Germany}

\date{\today}

\begin{document}

\maketitle

\begin{abstract}
The time dependent density matrix of a system with potential barrier
is studied using path integrals. The characterization of the initial state,
 which is
 assumed to be restricted to one side  of the barrier, and the time
 evolution of the density matrix lead to a three-fold path integral which
is evaluated in the semiclassical limit. The semiclassical trajectories
are found to move in the complex coordinate plane and barrier penetration
only arises due to fluctuations.
Both the form of the semiclassical paths and the relevant fluctuations change
significantly as a function of temperature. The semiclassical analysis leads to
a detailed picture of barrier penetration in the real time domain
and the changeover from thermal activation to quantum tunneling.
Deep tunneling is associated with quasi-zero modes in the fluctuation
spectrum about the semiclassical orbits in the long
time limit. The connection between this real time description of
tunneling and the standard imaginary time instanton approach
is established. Specific results are given for a double well potential
and an Eckart barrier.
\end{abstract}

\section{Introduction}

Semiclassical theories have been found extremely powerful in
understanding the dynamics of complex quantum mechanical systems.
 Special attention has been 
paid to theories of tunneling processes as they occur in
physics, chemistry, and biology. Currently, a variety of
quantum rate theories are in use explaining experimental
findings for several situations of interest \cite{hanggi}. Thereby, roughly
speaking, two 
different strategies
can be distinguished. The first  class of
approaches constructs the rate from purely thermodynamic
considerations. An example is the 
bounce or instanton method (also called Im$F$ method), originated by
Langer \cite{langer} and extended by several authors \cite{olschowski}.
In essence, 
tunneling rates are derived from the imaginary time dynamics in the inverted
potential. Other approaches of this type \cite{millerqtst,hontscha} start from
periodic orbit theory \cite{gutzwiller} in imaginary time. Tractable
 rate formulae are
obtained with the 
centroid method \cite{cdm} leading to a semi-empirical separation of
 dynamical and thermal 
factors. These methods are computationally very efficient and have
been applied successfully to systems as diverse as tunneling centers in
metals, Josephson junctions, or hydrogen bonds, to name but a few.
However,  closer examination reveals that these theories are based in one
 way or
another
 on {\it ad hoc} assumptions that are not derived from first
principles. For instance, the Im$F$ method postulates
a relation between the decay rate and the imaginary part of the
free energy. In fact, in some cases thermodynamic methods fail to 
predict the correct rate, e.g. they do not reproduce the energy
diffusion limited decay for very weakly damped systems at finite temperature.
Moreover, these methods are designed to describe incoherent decay
only. 

The second class of theories describes barrier crossing in terms of
dynamical quantities. Perhaps most familiar is Yamamoto's rate
formula \cite{yamamoto}, in essence a Kubo type formula 
relating the rate with a
flux-flux correlation function. As shown in \cite{tromp} it is  exact only
for scattering problems, while in multi- or metastable systems one
has to assume the  
existence of a plateau region for times long compared to typical
relaxation times but small compared to decay times. This restricts the
approach to incoherent decay.
 Recent advances in
the real time 
description of barrier crossing based on flux-flux correlations were
made e.g.\ by Voth {\it et al.} \cite{VCM89}, who incorporated 
analytically known dynamical factors into the  rate
expressions. Quite recently, Pollak and coworkers \cite{pollak,pollak1a} were
able to improve this 
idea by using a thermally symmetrized flux operator. A different way
of including dynamical information in an approximate way, favored
 by Miller and coworkers, employs semiclassical Initial Value
Representation (IVR) for the quantum propagator
\cite{miller}. Although  
quite successful at high to moderate temperatures, where the quantum
dynamics is 
governed by  quasi-classical above-barrier processes, these approaches
usually fail at low
temperatures where deep tunneling prevails.
Finally, we mention a kind of hybrid approach, the ``real
time'' instanton theory \cite{instanton} which includes tunneling
 in 
the real time propagator by means of instantaneous tunneling
transitions.  This method is 
restricted to multi-stable systems in the low temperature limit
\cite{weissbuch}. 
Hence, although the semiclassical theory of quantum
tunneling is often regarded to be well settled, this turns out to be
true only for some limiting cases.  What would be desirable  is a
semiclassical theory starting from first principles that covers the
entire range of
temperatures as well as coherent and incoherent tunneling processes.

In the realm of classical physics the theory of
thermally activated rates is rather firmly based. In
a seminal paper \cite{kramers}  Kramers determined
 thermal decay rates  from the equation of
motion for the phase space distribution function, i.e., from the real
time dynamics of the system. A corresponding treatment of tunneling in
the semiclassical limit seems not to be possible, since all real time
minimal action paths connecting two sides of the barrier have energies
{\it larger} than the barrier energy. In the dynamical approaches
discussed above these trajectories do account for tunneling
corrections to classical rates \cite{miller,kesha}, but it is usually argued
\cite{grossmann,heller} that within a 
semiclassical theory deep tunneling can only be described by
incorporating in addition imaginary time trajectories as they are used in
thermodynamic methods. The lack of a first principles
semiclassical theory of tunneling is intimately connected with a fundamental
shortcoming of  semiclassical real time propagators derived from a
single dominant path. Since tunneling
arises from coherent 
interference of waves, a satisfactory semiclassical
theory needs to capture this interference pattern in terms of
an appropriate family of real time paths rather than making {\it
ad-hoc} modifications of simple semiclassical propagators.

Very recently, we have proposed  a theory for transport 
across a barrier based solely on the real time dynamics of the density
matrix and the semiclassical approximation\cite{epl}.   As a notable
feature, the method applies 
equally well to damped and undamped systems and comprises in a unified
way the entire
range from thermally activated decay to low temperature tunneling. In
appropriate limits the results of other methods  are
recovered. Here, we explain the technicalities of the approach and
evaluate the semiclassical propagator in detail 
 for two paradigmatic
systems, namely,  a double well potential and an Eckart 
barrier. Thereby, we mainly concentrate on one-dimensional models. It
turns out that the corresponding theory already reveals the basic
structure and that the generalization to multidimensional systems,
though tedious in detail, is straightforward within the path integral
formalism \cite{dyn1}. 

The article is organized as follows. Next, in  Sec.~\ref{general}, we
outline the general semiclassical theory which is then used
 in Sec.~\ref{parabolic} to derive as a simple example the stationary
flux across a 
parabolic barrier. The main part of the paper studies the real time
dynamics from high down to vanishing temperature for the cases of a
double well potential (Sec.~\ref{double}) and an Eckart barrier
(Sec.~\ref{eckart}). Finally, in Sec.~\ref{conclu} we summarize the main
features of the approach and present our conclusions.

\section{General theory}\label{general}

We consider a statistical ensemble of quantum mechanical particles of
mass $M$ moving in a barrier
potential $V(q)$ at inverse temperature $\beta=1/k_{\rm B} T$. We
chose the coordinate $q$ so that the barrier top is located at $q=0$
and measure energies relative to the barrier energy by putting
$V(0)=0$. The  initial
nonequilibrium state is assumed to be of the form of an equilibrium
state restricted  to the left side
of the barrier. Below we will invoke the semiclassical approximation
which is appropriate provided the barrier
height $V_b$ is by
far the largest energy scale in the system. 
The
time evolution of the density matrix $\rho(t)=\exp(-i H
t/\hbar) \rho(0) \exp(i H t/\hbar)$ reads in coordinate representation
\begin{equation}
\rho(q_f,q_f',t) = \int\! dq_i dq_i'  G_t(q_f,q_i)\rho(q_i,q_i',0)
  G_t(q_f',q_i')^*\label{gt1}
\end{equation}
where  the real-time propagator is given by
\begin{equation}
G_t(q,q')=\langle q|\exp(-i Ht/\hbar)|q'\rangle\label{gt2}
\end{equation}
and $\rho(q_i,q_i',0)$ describes the initial state.
In principle, for our purpose  any initial distribution that matches onto
equilibrium on the left side and vanishes on the right side of
the barrier top is appropriate. As long as the
restricted equilibrium state gives vanishing probability to find the
particle on the right side of the barrier
top,  different initial
preparations lead to the same long time
behavior of the density 
matrix. Here we put explicitly
\begin{equation}
\rho(q,q',0)= Z^{-1} \rho_\beta(q,q') \theta(-q) \theta(-q')\label{gt3}
\end{equation}
for convenience with the proper normalization factor $Z$ and the
equilibrium density matrix
\begin{equation}
\rho_\beta(q,q')=\langle q|\exp(-\beta
H)|q'\rangle.\label{gt4}
\end{equation}

Now, employing the path integral representation for $\exp(\pm i t
H/\hbar)$ and $\exp(-\beta H)$, respectively, the above integrand in
Eq.~(\ref{gt1}) can
be written as a 
three-fold path integral  where two real time paths $q(u)$ and
$q'(u)$ run in the interval $0\leq u\leq t$ from  
$q_i$ and $q_i'$ to fixed endpoints $q_f$ and $q_f'$, respectively,
while those former coordinates
are connected by an imaginary time path $\bar{q}(\sigma)$ in the
interval $0\leq\sigma\leq\hbar\beta$, see fig.~\ref{loop}. The real
time paths describe the 
time evolution of the system and the imaginary time path the initial
state. Of course, a more complete theory would explicitly include the
coupling to a heat bath environment. In fact, the general scheme of
this approach in the case of 
damped systems  has
already been given elsewhere \cite{dyn1}. Much of the analysis
presented below can 
in principle be extended to this situation, however, only a limited
number of steps can be carried out analytically due to
the more complicated form of the effective action functionals. Here,
we limit ourselves to undamped motion 
which allows to treat deep tunneling without resorting to numerical
methods. This way, the guiding concepts will become more transparent.

The density matrix $\rho(q,q',t)$ contains all
information about the nonequilibrium quantum process, in particular, 
the average of the operator $F=[p \delta(q)+\delta(q) p]/2 M$
gives the flux out of the metastable state,
i.e., in coordinate representation
\begin{equation}
J(t)=(\hbar/2 i M)\, [\partial\rho(q_f,-q_f,t)/\partial q_f]_{q_f=0}
.\label{gt5}
\end{equation}
If  the flux becomes quasi-stationary,  $J(t)= J_{\rm
fl}$  within a certain ``plateau
region'' of time, the escape rate follows from $\Gamma=J_{\rm fl}$.

While  an exact solution of Eq.~(\ref{gt1}) for anharmonic
barrier potentials is not possible, a high barrier naturally suggests a
semiclassical approximation.
In the  semiclassical limit the above path integrals are
dominated by minimal action paths determined by Hamilton's equation of
motion either in the potential $V(q)$ (for the real time propagators)
or $-V(q)$ (for the equilibrium density matrix). 
Each path contributes with an exponential factor containing its
minimal action and a prefactor  
arising from the Gaussian fluctuations about the minimal action paths.
Specifically, the action in real time reads 
\begin{equation}
S(q,q')=\int_0^t du [M \dot{q}^2/2-V(q)]\label{gt6}
\end{equation}
while its imaginary time version, the so-called Euclidian
action,  is given by
\begin{equation}
 \bar{S}(q,q')=\int_0^{\hbar\beta}d\sigma
[M\dot{\bar{q}}^2/2+V(\bar{q})].\label{gt7}
\end{equation} 
Thus, in the Gaussian semiclassics the propagator
(\ref{gt2}) is approximated as
\begin{equation}
G_t(q,q')=\sum_{{\rm cl.paths}}\ \sqrt{A(q,q')} \,
\exp\left[\frac{i}{\hbar} S(q,q')-i\frac{\pi}{2}
\nu\right]\label{gt6a}
\end{equation}
where $A(q,q')= [-\partial^2 S(q,q')/\partial q\partial q']/2\pi i
\hbar$ and $\nu$ 
is the Maslov index. Throughout this paper we also use an equivalent
representation of the prefactor, namely, 
\begin{equation}
A(q,q') = \frac{i M}{{2\pi\hbar}} \, \left[ \dot{q}(0)
\dot{q}(t) \frac{\partial^2 W(q, q')}{\partial E^2}\right]^{-1}\label{pre}
\end{equation}
where $W(q,q')=\int_{q}^{q'}dq'' p = S(q,q')+ E t$ is the short action. 
The corresponding
approximation to the equilibrium density matrix (\ref{gt4})
follows by formal analytic continuation $t\to -i\hbar\beta$, i.e.,  by
replacing  $S(q,q')$ by 
$i\bar{S}(q,q')$ in Eq.~(\ref{gt6a}) with $\nu=0$.
As a result, the integrand in
Eq.~(\ref{gt1}) is completely determined by  classical mechanics
in real and imaginary
time, respectively,  and dominated by an action factor
$\exp[-\Sigma(q_f,q_f'|q_i,q_i')/\hbar-i\pi(\nu-\nu')/2]$ with
\begin{equation}
\Sigma(q_f,q_f'|q_i,q_i') = -iS(q_f,q_i)+\bar{S}(q_i,q_i')+i
S(q_f',q_i').\label{gt8}
\end{equation}

With the approximate integrand at hand, it is consistent
to evaluate the ordinary
integrations in Eq.~(\ref{gt1}) in stationary phase. 
The stationary phase points are determined by minimizing  $\Sigma$
with respect to the initial coordinates $q_i,q_i'$,
i.e.,
\begin{equation}
 \left.\frac{\partial\Sigma}{\partial q_i}\right|_{(q_f,q'_f)}=0,\ \
\ \ \left.\frac{\partial\Sigma}{\partial 
q_i'}\right|_{(q_f,q'_f)}=0.\label{gt9}
\end{equation}
Since the endpoints $q_f, q_f'$ are fixed, the resulting stationary
phase points
$q_s(t)$ and $q_s'(t)$ are functions of time with
$q_s(0)=q_f$, $q'_s(0)=q_f'$.
For finite $t$ these roots  are in general complex.  The dominant 
imaginary time path $\bar{q}_s(\sigma)\,$ connects $q_s'(t)$ with
$q_s(t)$, and the two real time paths $q(u)$ and $q'(u)$ connect
$q_s(t)$ and $q_s'(t)$ with $q_f$ and $q_f'$, respectively. Hence, the
steepest--descent approximation naturally provides a mapping from
the integration contour in the complex time plane onto a loop in the
complex coordinate space connecting the endpoints [fig.~\ref{loop}]. 
To avoid potential
 confusion with other  methods,  we emphasize that
the appearance of complex paths has 
nothing to do 
with tunneling but rather is merely a consequence of the stationary
phase approximation and
holds also for systems with no barrier at all. In fact, it turns out
that  the complex semiclassical real-time trajectories used here never
cross the barrier top, in contrast to paths emerging from {\it ad hoc}
complexification procedures occasionally adopted to describe barrier
 penetration
\cite{miller74}.  

Starting from the steepest descent conditions (\ref{gt9}) and exploiting
Hamilton-Jacobi mechanics, one immediately derives 
\begin{equation}
p_s(0)=i \bar{p}_s(\hbar\beta)\  ,\  p_s'(0)=i \bar{p}_s(0)\ ,
\ E=E'=\bar{E}.\label{gt10}
\end{equation}
Here, $p_s(u)\,\, [p_s'(u)]$ is the momentum of
the real time path $q(u)\,\, [q'(u)]$ with energy $E\,\, [E']$
connecting $q_s\,\, [q_s']$ 
and $q_f\,\,[q_f']$; accordingly,
$\bar{p}_s(\sigma)$  denotes the momentum of  the imaginary time path
$\bar{q}_s(\sigma)$ running from $q_s'$ to $q_s$ 
with Euclidian energy $\bar{E}=-\bar{p}_s^2/2M+V(\bar{q})$, see
fig.~\ref{loop}.
Eq.~(\ref{gt10})
can  also be expressed as
$d\Sigma/dt=0$ with the solution
\begin{equation}
\Sigma(q_f,q_f'|q_s,q_s')=\bar{S}(q_f,q_f').
\label{gt11}
\end{equation}
Hence along the loop of steepest descent paths the full action is just
 given by the equilibrium action and thus independent of time. 
 Differentiating Eq.~(\ref{gt11}) with respect to
$q_f, q_f'$, one finds 
\begin{equation}
p_s(t)=i\bar{p}_0(\hbar\beta),\ \ \  p_s'(t)=i
\bar{p}_0(0)\label{gt12}
\end{equation}
 where $\bar{p}_0(\sigma)$ is now the momentum of
the imaginary time path $\bar{q}_0(\sigma)$ connecting $q_f'$ with
$q_f$ in imaginary time $\hbar\beta$. This path has Euclidean energy
${\bar E}_f$ which depends on $q_f, q_f'$ and $\hbar\beta$ but not on
$t$. Hence, we first deduce that the energies in 
Eq.~(\ref{gt10}) are given by $\bar{E}_f$, which implies energy and
momentum conservation throughout the loop in fig.~\ref{loop}. Secondly, we
arrive at the
remarkable result that the sequence of time-dependent stationary phase
points $q_s(t)\,\, 
[q_s'(t)]$ is itself a minimal action path starting at   
$q_s(0)=q_f\,\,
[q_s'(0)=q_f']$ with energy $\bar{E}_f$.

To complete the ordinary integrations in Eq.~(\ref{gt1}) 
over the initial coordinates $q_i, q_i'$ we transform to fluctuations
$y=q_i-q_s$ and $y'=q_i'-q_s'$ about the stationary phase
points. An expansion of the full action 
(\ref{gt8}) for fixed endpoints $q_f, q_f'$ around the stationary phase
points up to second order leads to 
$\Sigma(q_f,q_f'|q_i,q_i') = \bar{S}(q_f,q_f')+
\delta\Sigma^{(2)}(y,y')$ with
\begin{equation}
\delta^{(2)}\Sigma(y,y')= \frac{1}{2}\  (y,y')\ \Sigma^{(2)}\ \left(
\begin{array}{c}
y\\
y'
\end{array}\right)\label{gt13}
\end{equation}
where 
\begin{equation}
\Sigma^{(2)}= \left(
\begin{array}{c c}
\Sigma_{ss} \Sigma_{ss'}\\
\Sigma_{ss'} \Sigma_{s's'}
\end{array}\right)\label{gt13a}
\end{equation}
is the matrix of second order derivatives, 
$\Sigma_{ss}=\partial^2\Sigma(q_i,q_i')/\partial q_i^2$ etc.,
to be taken at $q_i=q_s, q_i'=q_s'$. 

Inserting Eq.~(\ref{gt3}) into Eq.~(\ref{gt1}),
the integrand now reduces to a product of
Gaussian weight factors for deviations from the stationary phase
points and an initial state factor $\theta(-q_s-y)\theta(-q_s'-y')$
describing deviations from thermal equilibrium at 
$t=0$. Provided there is only one semiclassical  path for each of the
propagators, we 
obtain from Eq.~(\ref{gt1}) by virtue of Eqs.~(\ref{gt11}) and 
(\ref{gt13}) the semiclassical time dependent density
matrix in the form
\begin{equation}
\rho(q_f,q_f',t)=\frac{1}{Z}\rho_\beta(q_f,q_f')\, g(q_f,q_f',t).\label{gt17}
\end{equation}
Here, deviations from equilibrium are described by  a ``form factor''
\begin{equation}
g(q_f,q_f',t)= \frac{1}{\pi} \int_{-\infty}^{u(q_s)} dz\, 
\int_{-\infty}^{u'(z,q'_s)} dz'\, {\rm e}^{-(z^2+{z'}^2)},\label{gt18}
\end{equation}
where 
\begin{equation}
u(q_s)= -q_s\, \sqrt{\frac{{\rm
Det}[\Sigma^{(2)}]}{2\hbar\Sigma_{s's'}}},\ \ \ \ u'(q_s',z)= -q_s'
\sqrt{\frac{\Sigma_{s's'}}{2\hbar}} + z \frac{\Sigma_{ss'}}{\sqrt{{\rm
Det}[\Sigma^{(2)}]}}\label{gt15}
\end{equation}
with Det$[\Sigma^{(2)}]=\Sigma_{ss}\Sigma_{s's'}-(\Sigma_{ss'})^2$.
In deriving Eq.~(\ref{gt17}) we invoked that Hamilton Jacobi
mechanics implies \cite{berry}  
\begin{equation}
\left[\frac{A(q_f,q_s) \bar{A}(q_s,q_s') A(q_f',q_s')}{{\rm
Det}[\Sigma^{(2)}]}\right]^{1/2}=\bar{A}(q_f,q_f').\label{gt16}
\end{equation}
Note that for an initial equilibrium state, formally
$\theta(\cdot)\to 1$ in Eq.~(\ref{gt3}) so that $u, u'\to
 \infty$ in Eq.~(\ref{gt18}), the form factor becomes 1 and the semiclassical
 density matrix is in fact  stationary.
If there is more than one classical path one has to sum in
Eq.~(\ref{gt17}) over the contributions of all of them.
Certainly, the above formulae (\ref{gt18}) and (\ref{gt15}) are
only applicable as long as the Gaussian semiclassical and stationary phase
approximations are valid, i.e.\ as fluctuations are sufficiently
small. This will be seen to be no longer the case for low 
temperatures and/or very long times. How the classical paths in the
complex plane can then be
used as a skeleton for an 
extended semiclassical/stationary phase calculation will be shown
below. 

In the remaining parts of the article we apply the general formalism
 to specific barrier potentials. Thereby,  since we are
particularly interested in the flux  across the
barrier, 
 we restrict our investigation to non-diagonal end-coordinates
$q_f$ and $q_f'=-q_f$ close to the barrier top. This does not mean
 that we may constrain ourselves to study only local dynamics near the
 barrier top. Especially for
lower temperatures,   the nonequilibrium state in the barrier region
 is predominantly governed  by global properties of the potential.

\section{Parabolic barrier}\label{parabolic}

The semiclassical and the stationary phase approximations are always exact
for quadratic potentials. Hence, as a simple test case we consider
first a parabolic barrier
\begin{equation}
V(q)= - \frac{1}{2} M \omega_b^2\, q^2.\label{para1}
\end{equation}
Accordingly, the imaginary time dynamics runs in a harmonic oscillator
potential. For the minimal action path $\bar{q}_0(\sigma)$ connecting
 $-q_f$ with $q_f$ in time $\hbar\beta$ one obtains
\begin{equation}
\bar{q}_0(\sigma)= \frac{q_f}{\sin(\omega_b\hbar\beta/2)}\,
\sin\left[\omega_b(\sigma-\hbar\beta/2)\right].\label{para2}
\end{equation}
This leads to the well-known equilibrium density matrix
\begin{equation}
\rho_\beta(q_f,-q_f)=
\frac{1}{\sqrt{4\pi\delta_b^2\sin(\omega_b\hbar\beta)}}\  
\exp\left[ - \cot(\omega_b\hbar\beta/2)\,
\frac{q_f^2}{2\delta_b^2}\right]\label{para3}
\end{equation}
with the relevant length scale $\delta_b=\sqrt{\hbar/2M\omega_b}$.

The real-time dynamics simply follows. The classical real-time paths
$q(u)$ and $q'(u)$ lead to the endpoints $q_f$ and $-q_f$,
respectively, and hence obey $q(t)=q_f$, $q'(t)=-q_f$. On the other
hand, the stationary phase condition (\ref{gt10}) implies
$\dot{q}(t)=i\dot{\bar{q}}(\hbar\beta)$,
$\dot{q}'(t)=i\dot{\bar{q}}(0)$ and we readily find
$q(u)=\bar{q}(\hbar\beta-i t+i u)$, i.e.,
\begin{eqnarray}
q(u)&=&\frac{q_f}{\sin(\omega_b\hbar\beta/2)}\,
\sin\left[\omega_b(\hbar\beta/2-it+iu)\right],\nonumber\\
q'(u)&=&q(u+i\hbar\beta),\ \ 0\leq u\leq t. \label{para4}
\end{eqnarray}
At time $t$ the
imaginary time path 
$\bar{q}_0(\sigma)$ from $-q_f$ to $q_f$  is mapped onto the path
$\bar{q}_s(\sigma)=q(i\hbar\beta-i\sigma), 0\leq \sigma\leq
\hbar\beta$ connecting 
$q_s'(t)=q'(0)$ with $q_s(t)=q(0)\, $ (fig.~\ref{hightc}). 
The stationary phase points  
\begin{eqnarray}
q_s(t)&=&\frac{q_f}{\sin(\omega_b\hbar\beta/2)}
\sin\left[\omega_b(\hbar\beta/2-it)\right]\nonumber\\
q_s'(t)&=& -q_s(t)^*\label{para4aa}
\end{eqnarray} 
 are as functions of $t$ also classical paths moving away from the
barrier top as $t$ increases -- $q_s(t)$ to the 
right and $q_s'(t)$ to the left for $q_f>0$. For 
longer times $\omega_b t\gg 1$, the stationary phase points
asymptotically tend towards the 
limiting trajectories starting from $q_f=0$ referred to as asymptotes
henceforth. 
Similar to separatrices in classical phase space, these
asymptotes divide the complex plane in  regions of negative and
positive Euclidian energy: in the sectors including  the real
axis classical real time motion has 
$\bar{E}<0$ (but $E\leq 0$ or $E>0$)  while in the remaining parts 
$\bar{E}>0$.  However, in 
contrast to simple classical separatrices the asymptotes are temperature
dependent. The  angel $\alpha$ of the $q_s$ and $q_s'$-asymptotes with the
positive and  negative real axis, respectively, is found as  
\begin{equation}
\alpha=\frac{\pi-\omega_b\hbar\beta}{2}.\label{para4b}
\end{equation}

Now, for the nonequilibrium preparation (\ref{gt3}) the initial
 coordinates $q_i, q_i'$
are constrained to Re$\{q_i\}$, Re$\{q_i'\}\leq 0$. Since $q_s(t)$ and
$q_s'(t)$ are on different sides of the barrier, $\rho(q_f,-q_f,t)$ gains
nonvanishing values {\it only due to fluctuations} that effectively
shift $q_i$ away from $q_s$ and across the barrier top [see
Eq.~(\ref{gt15})]. For the parabolic barrier potential the matrix
elements in (\ref{gt13a}) take the simple form
\begin{eqnarray}
\Sigma_{ss}&=& M \omega_b\left[\cot(\omega_b\hbar\beta)-i\coth(\omega_b
t)\right],\ \ \Sigma_{s's'}=\Sigma_{ss}^*,  \nonumber\\
\Sigma_{ss'}&=&-\frac{M\omega_b}{\sin(\omega_b\hbar\beta)}\label{para4c}
\end{eqnarray}
so that the matrix $\Sigma^{(2)}$ can easily be diagonalized. One
finds for  the eigenvalues  
\begin{equation}
\frac{\lambda_{\pm}}{M\omega_b}=\cot(\omega_b \hbar\beta)\pm
\left[\cot(\omega_b\hbar\beta)^2-\frac{1}{\sinh(\omega_b t)^2}\right]^{1/2}.
\label{para4d}
\end{equation}
While, in principle,  with Eq.~(\ref{gt18}) we can now evaluate
the complete dynamics of the density matrix, we will concentrate here on
the long time asymptotics of the nonequilibrium state. 
Then, in the asymptotic region 
$\omega_b t\gg 1$ the eigenvalue
$\lambda_-$ tends to zero as $\lambda_-\propto M\omega_b\exp(-\omega_b t)$,
 reflecting the instability of the parabolic barrier. Hence,
fluctuations around the stationary phase points with the
least action increase occur in the direction of the eigenvector with eigenvalue $\lambda_-$. These fluctuations are of the form 
 $y_i=|y_i| \exp[i(\alpha+\omega_b\hbar\beta)]$ and $y_i'=|y_i|
\exp(i\alpha)$ so that  $q_i$
and $q_i'$  move 
simultaneously along their asymptotes meeting at the barrier
top (see fig.~\ref{hightc}). Now, inserting the matrix elements
(\ref{para4c}) and 
the stationary phase points (\ref{para4aa}) into Eq.~(\ref{gt15}) and
considering the limit $\omega_b t\gg 1$, the
relevant form factor turns out to be stationary
\begin{equation}
g_{\rm fl}(q,-q)= \frac{1}{\sqrt{\pi}}\ \int_{-\infty}^{i 2 q \Omega} dx
\ {\rm e}^{-x^2}\label{para6}
\end{equation}
with $\Omega=\sqrt{\cot(\omega_b\hbar\beta/2)/(8\delta_b^2)}$. 
The corresponding
constant flux across the barrier is  obtained
from Eqs.~(\ref{gt5}) and (\ref{gt17}) as  
\begin{equation}
J_{\rm fl}=\frac{\hbar}{2 Z M} \rho_\beta(0,0)\ \left. \frac{\partial
g_{\rm fl}(q_f,-q_f)}{\partial 
q_f}\right|_{q_f=0}
\label{para5}
\end{equation}
which leads to the well-known result
\begin{equation}
\Gamma=J_{\rm fl}=  \frac{\omega_b}{4\pi}\, 
\frac{1}{Z \sin(\omega_b\hbar\beta/2)}. \label{para7}
\end{equation}
Here, $Z$ denotes an appropriate normalization constant which cannot
be derived from the pure parabolic potential. This is not really a
problem since  realistic potentials always exhibit a well-behaved potential
minimum and then $Z$ follows  e.g.\ as the relative normalization with
respect to this minimum.
Note that for the quadratic potential the results (\ref{para6}) and
(\ref{para7}) are formally valid for all
times $\omega_b t\gg 1$. 
 However, it was shown in \cite{paper0}
that due to the lack of a well-behaved ground state it makes
physically only sense to use the parabolic barrier in 
$T>T_c$ where $\omega_b \hbar/k_{\rm B} T_c=\pi$. For lower temperatures 
$\omega_b\hbar\beta\to \pi$ large quantum fluctuations render the
Gaussian approximation insufficient.  Interestingly, the
rate expression 
(\ref{para7}) diverges at the lower temperature $T_0=T_c/2$ only where
the  parabolic density matrix (\ref{para3}) ceases to exist.

\section{Double well potential}\label{double}

A model well-behaved for the entire range of temperatures with many
applications is the bistable dynamics of a particle moving in 
 a double well potential
\begin{equation}
V(q)=-\frac{M\omega_b^2}{2}\, q^2\, \left[ 1-
\frac{q^2}{2 q_a^2}\right]\label{db1}
\end{equation}
Here, 
the  barrier is located at $q=0$, the wells at $q=\pm q_a$, and the
barrier height is $V_b=-V(q_a)=(M\omega_b^2/4) q_a^2$. 
This potential 
exhibits rich quantum dynamics, namely,  incoherent hopping between
the wells over
a broad range of temperatures  that changes to coherent oscillations for
$T\to 0$. Due to the complexity of the dynamics this is a highly
nontrivial problem for the semiclassical approach where the ratio
$\delta_b/q_a$ serves as the small parameter.

\subsection{Thermal equilibrium}\label{doubleequi}

The Euclidian mechanics in the inverted potential $-V(q)$ can be
solved exactly using Jacobian elliptic functions
\cite{abramowitz}. For the general 
solution one  obtains
\begin{equation}
\bar{q}_0(q_f,\sigma)= B\, {\rm
sn}\left[\omega(B) \sigma-\phi_f|m\right],\ \ \ 0\leq
\sigma\leq \hbar\beta\label{db2} 
\end{equation}
where the boundary conditions $\bar{q}_0(q_f,0)=-q_f$ and
$\bar{q}_0(q_f,\hbar\beta)=q_f$ fix the amplitude $B$ and phase $\phi_f$.
Since the potential is no longer purely quadratic -- depending on
temperature -- there may be several solutions
each of them with another amplitude.
In Eq.~(\ref{db2})  the
frequency is given by
\begin{equation}
\omega(B)= \omega_b\ \sqrt{1-\eta^2},\ \ \ \eta^2=\frac{B^2}{2
q_a^2},\label{db3}
\end{equation}
and the phase can be represented as an incomplete elliptic integral
\begin{equation}
 \phi_f=F(q_f/B|m)=\int_0^{q_f/B}dx
 \frac{1}{\sqrt{(1-x^2)(1-mx^2)}}\label{db4}
\end{equation}
with the so-called modul $m=\eta^2/(1-\eta^2)$. From the boundary
condition $\bar{q}_0(\hbar\beta)=-\bar{q}_0(0)$ and the periodicity
of the Jacobian function, ${\rm sn}[z+2 r K(m)|m]= (-1)^r {\rm sn}[z|m],
r=1,2,3,\ldots$ with 
$K(m)=F(1,m)$, the amplitude $B$ is determined by
\begin{equation}
\omega(B) \hbar\beta=2 r K(m)+[1+(-1)^{r}] \, \phi_f.\label{db4a} 
\end{equation}
Since $K(m), \phi_f>0$, for fixed $\omega_b\hbar\beta$ real solutions
to this equation exist only for a finite number of integers $r\geq 0$.

Let us briefly discuss the trajectories $\bar{q}_0(q_f,\sigma)$ as
the temperature is lowered. For high temperatures  only solutions of 
Eq.~(\ref{db4a}) with $r=0$ exist corresponding to direct paths from
$-q_f$ to $q_f$; particularly, $\bar{q}_0(0,\sigma)=0$. 
  As the
temperature drops below the critical temperature
\begin{equation}
T_c=\hbar\omega_b/\pi k_{\rm B},\label{db4aa}
\end{equation}
i.e.\ $\omega_b\hbar\beta>\pi$,
solutions of Eq.~(\ref{db4a}) with $r=1$ arise. Then, for $q_f=0$
 the
barrier top can be joined with itself also by two nonlocal paths
denoted by
$\bar{q}_{\pm}(0,\sigma)$ oscillating in $-V(q)$ to the right and to
the left with amplitudes $\pm q_1$, respectively, 
and  energy $\bar{E}_1=V(q_1)$. With further decreasing temperature
 $q_1$ grows and eventually saturates at $ q_a$ for $T\to 0$. For finite
$q_f$ the situation is rather similar: 
oscillating paths $\bar{q}_{\pm}(q_f,\sigma)$ exist for all
$q_f<q_1$. These paths connect $-q_f$ with $q_f$  via a
turning point at $\pm q_1$, thus, differing from
$\bar{q}_{\pm}(0,\sigma)$ only by a phase shift. 
The described  scenario repeats in an analog way  at
all $T=T_c/r, r=2,3,4,\ldots$, where $r$ counts the number of turning
points. At zero temperature all these
oscillating paths reach 
$\pm q_a$ with the same energy $\bar{E}_a=V(q_a)$ and are then
called instantons.
 
Now that all proper Euclidian trajectories are identified, the
semiclassical equilibrium state follows readily. For high
temperatures $T>T_c$ and end-coordinates $q_f$ near the barrier top,
$\rho_\beta(q_f,-q_f)$ basically coincides with the parabolic result
(\ref{para3}) and  anharmonic corrections are negligible. This
situation changes drastically for temperatures near $T_c$. Then, the
bifurcation of new classical paths leads to large quantum
fluctuations and one has to go beyond the Gaussian approximation of
the fluctuation integral. Slightly below $T_c$ a caustic appears for
$q_f=q_1$. Since 
the region around $T_c$ was 
already studied in detail elsewhere \cite{paper0} we omit here this
crossover region and proceed with 
temperatures sufficiently below $T_c$ where near the barrier top 
Gaussian semiclassics 
is again applicable. It turns
out that the paths newly emerging near 
$T_c$  are
stable and dominate $\rho_\beta(q,q')$ for all $T_c>T>0$ while the
unstable ``high temperature'' path and those springing up at lower $T$
give  negligible
contributions. Since for $q_f<q_1$  all  paths
$\bar{q}_{\pm}(q_f,\sigma)$ differ only
by a phase shift, one has for the corresponding actions
\begin{equation}
\bar{S}_{\pm}(q_f,-q_f)=\bar{S}_{+}(0,0)=\bar{S}_{-}(0,0)\label{db5aa}
\end{equation}
 so that
\begin{equation}
\rho_\beta(q_f,-q_f)= 2[\bar{A}(q_f,-q_f)]^{1/2} 
\exp[-\bar{S}_+(0,0)/\hbar],\ \ \ q_f<q_1.\label{db5ab}
\end{equation}
 Accordingly, the matrix element
$\rho_\beta(q_f,-q_f)$   changes to a non-Gaussian 
distribution with a  local minimum at $q_f=0$ and two maxima at
$q_f=\pm q_1$. Thereby $\bar{S}_+(0,0)<0$,  so that
the probability $\rho_\beta(0,0)$ to find the particle at $q=0$ is
substantially enhanced compared 
to its classical value. 

For $T\to 0$ it is no longer sufficient to include only the
 trajectories with $r=1$ in the semiclassical analysis but rather all
 other paths with $r>1$ must also be taken into account. This is due
 to the fact that the smaller action factors of these latter paths are
 compensated for by zero mode phase factors from the corresponding
 fluctuation path integrals. Accordingly,  
 all instanton contributions are
 summed up to yield e.g.\ for coordinates near the barrier top
\begin{equation}
\rho_\beta(q_f,-q_f)= \frac{8}{\delta_a\sqrt{2\pi}}\ {\rm e}^{-\beta
[V(q_a)+\hbar\omega_a/2]-\bar{W}_a/\hbar} \left[
\cosh\left(\frac{\hbar\beta\Delta}{2}\right)
+\cosh\left(\frac{q_f q_a}{2\delta_a^2}\right)
\sinh\left(\frac{\hbar\beta\Delta}{2}\right)\right]. 
\label{db6}
\end{equation}
Here,  $\bar{W}_a\equiv\bar{W}(-q_a,q_a)=-\hbar\beta
\bar{E}+\bar{S}(-q_a,q_a)$ is the 
short action for an instanton from $-q_a$ to $q_a$. Further, 
\begin{equation}
\Delta= \omega_a \, \frac{4 q_a}{\sqrt{2\pi} \delta_a}\
\exp(-\bar{W}_a/\hbar) \label{db6b}
\end{equation}
 denotes the  WKB tunnel splitting
with the well frequency $\omega_a=\omega_b\sqrt{2}$ and
$\delta_a^2=\hbar/2M\omega_a$.

\subsection{Dynamics of stationary phase points}\label{doublestat}
 
As in case of the parabolic barrier, the stationary real-time paths can
be directly infered from the Euclidian dynamics at $t=0$. From
Eq.~(\ref{db2}) and the stationary phase condition we have 
\begin{equation}
q_s(t)= B\, {\rm sn}[\phi_f-i\omega(B) t|m],\ \ \ \
q_s'(t)=-q_s[(-1)^{r+1}t]\label{db7}
\end{equation}
and $\bar{q}_0(\sigma)$ is mapped at time $t$ onto
$\bar{q}_s(\sigma)=\bar{q}_0[\sigma+i(-1)^{r+1}t]$ where $r$ follows
from Eq.~(\ref{db4a}). 
In the sequel we always formulate the semiclassical theory in terms of
the real time paths $q_s, q_s'$ that ``start''
 at the endpoints $q_f, -q_f$, respectively, and reach the initial
points $q_i, q_i'$ after time $t$. Since the endpoints
$q_f,-q_f$ are fixed, while the most relevant initial coordinates
depend on time, this backward view of the dynamics is in fact more
transparent. The real time trajectories now start from the end
coordinates we are interested in and lead to the relevant initial
coordinates that  need to
be integrated over with an integrand weighted according to the initial
deviations from equilibrium.
The  path $q_s(t)$ runs in the complex
coordinate plane as a periodic orbit with period
$t_p(q_f)= 2  K(1-m)/\omega(B)\, $ (fig.~\ref{fluczero}). Within one
period it connects $q_f$ with $q_f$ via a loop  crossing the
real axis also after time $ t_p(q_f)/2$ at the point
$q_c(q_f)=q_s[q_f,t_p(q_f)/2]\geq q_a$. Thus $q_s(t)$  stays always on the
same side of the barrier top and likewise $q_s'(t)$ on the other side so that
 the complex dynamics of the stationary
phase points starting from $q_f$ and $-q_f$, respectively, reflects a
bounded motion in either of the potential wells. 

Let us consider the stationary orbits as the temperature decreases. For high
temperatures $T>T_c$, i.e.\ $r=0$,  each $q_f$-dependent loop carries
its own period 
$t_p(q_f)$ and energy $E(q_f)$. If $q_f\neq 0$, $t_p$ is small for
$T\gg T_c$ and the real time dynamics corresponds to a fast 
bouncing back and forth  in the well. As the temperature
is lowered the period grows while simultaneously the ``width'' of the
loop $q_c(q_f)$ shrinks. In the special case $q_f=0$  the real
time path reduces to a constant $q_s(0,t)=0$.
For temperatures $T<T_c$ the situation changes according to the
appearence  of new oscillating Euclidian paths
$\bar{q}_{\pm}(q_f,\sigma)$ for $q_f<q_1$.
In contrast to the high temperature case {\it all} stationary phase point
paths with $q_f<q_1$  have then the same period $t_p(q_f)=t_p(q_1)$
and energy $E(q_f)=\bar{E}_1$, and differ only in their respective phases.
Special cases are $q_f=0$ and $q_f=q_1$: The path
$q_s(q_f,t), q_f\to 0$ 
runs along the imaginary axis, while  the orbit $q_s(q_1,t)$
degenerates to a usual well oscillation along the real axis.
These properties have a direct effect on
the corresponding 
actions. One finds by employing Cauchy's theorem that after each period 
\begin{equation}
S[q_s(q_f, n t_p),q_f]=S[q_s(q_1,n t_p),q_1], \ \ \ q_f\leq q_1.\label{db7a}
\end{equation}
Hence,  all $q_s(q_f,t)$ for
$q_f<q_1$ can be seen as phase shifted copies of the specific real
 path $q_s(q_1,t)$ having the same energy, period, and action
increase during  one period.
In particular, the period $t_p(q_1)$ is large for $T{\textstyle
 {\lower 2pt \hbox{$<$} \atop \raise 1pt \hbox{$ \sim$} }}T_c$ when
$q_1$ is still small, $t_p(q_1)\approx \ln(q_a/q_1)/\omega_b$, and
drops down to 
$t_p(q_a)=2\pi/\omega_a$ in the limit  $T\to 0$.

\subsection{Nonequilibrium dynamics for high and moderately low
temperatures}\label{nondb} 

For $t=0$ the density matrix is given by the initial state
(\ref{gt3}). The
semiclassical time
evolution of this state follows
by inserting the proper classical paths into Eq.(\ref{gt17}).
In the sequel, we mainly focus on the long time dynamics and are especially
interested in a plateau region where the time evolution becomes
quasi-stationary. 

We start by addressing the question when a plateau
region does exist at all.  Inserting the classical paths into
$g(q_f,-q_f,t)$ in Eq.~(\ref{gt15}), 
the detailed analysis reveals that this function becomes stationary when
the ratio $q_s(t)/p_s(t)$ reduces to a constant. Since this will only
occur within the parabolic barrier region, a least upper bound
for a plateau region 
follows from the time interval within which a particle starting at a
typical point $q_b$ 
near the barrier top continues to 
experience a nearly parabolic potential. This leads to  $t\ll
t_p(q_b)$. The lower bound is obvious: 
it is given by the transient time near the top, i.e.\ by
$1/\omega_b$. Hence, a plateau region can be estimated
to occur as long as a there is  clear separation of time scales
between local barrier 
motion and  global well oscillations, i.e., $1\ll \omega_b t\ll
\omega_b t_p(q_b)$. Particularly, in the 
temperature domain where a plateau region exists,  the
Gaussian/stationary 
phase approximations are valid and we can
actually calculate a rate. These 
approximations break down  when
the stationary phase  
points move far from the barrier top and times of order $t_p(q_b)$
become relevant for the barrier crossing.
This range will be addressed in
the next section.

For high temperatures $T>T_c$ the typical length scale $q_b$ can be
identified with $\delta_b=\sqrt{\hbar/2M\omega_b}$. Then, the separation
of time scales fails for very high temperatures where
$k_{\rm B} T {\textstyle
 {\lower 2pt \hbox{$>$} \atop \raise 1pt \hbox{$ \sim$} }}
8 V_b$ meaning that the thermal energy is  of the same order or larger
than the barrier height. With decreasing temperature $t_p$ grows so
that for $\omega_b\hbar\beta$ of order 1 a wide plateau range appears with
$t_p\approx \ln(q_a/\delta_b)/\omega_b$. In the corresponding density matrix
anharmonic corrections are small, and we obtain approximately the
parabolic result (\ref{para6}). To get the rate, here, the proper normalization
constant $Z$ is taken as 
 the partition function of the harmonic well oscillator
\begin{equation}
Z=\frac{1}{2 \sinh(\omega_a\hbar\beta/2)}\, {\rm e}^{\beta V_b}.\label{nondb1}
\end{equation}
Hence, from Eq.~(\ref{para5}) one regains the known result
\begin{equation}
\Gamma=\frac{\omega_b}{2\pi}\,
\frac{\sinh(\omega_a\hbar\beta/2)}{\sin(\omega_b\hbar\beta/2)}\, {\rm
e}^{-\beta V_b}\label{nondb2}
\end{equation}
with the exponential Arrhenius factor and a characteristic
$\hbar$-dependent prefactor 
that formally tends to $\omega_a/\omega_b$ in the classical
limit and describes the quantum enhancement of the rate as $T_c$ is approached.

At this point we have to be very careful: a
detailed analysis \cite{dyn1} of the full density matrix 
$\rho(q_f,q_f',t)$, not only
of its nondiagonal part,  reveals that the nonequilibrium effects
described by the flux state are restricted to the barrier region only
in the presence of damping, consistent with the fact that finite
temperature decay rates require coupling to a heat bath.
In the absence of damping the full density matrix does not become
quasistationary and the real time trajectories explore the strongly
anharmonic range of the potential.
 Hence, an evaluation
of the rate based upon  a supposedly quasi-stationary flux state
$\rho_{\rm fl}(q_f,-q_f)$ for the 
undamped case corresponds to the transition state theory
result. We refer to \cite{dyn1} for a detailed discussion of this point.

As the temperature reaches $T_c$ large quantum fluctuations occur and
the impact of anharmonicities becomes substantial. A detailed study of
the bifurcation range around $T_c$  is quite  tedious and was already
presented in \cite{dyn2}. Thus, we omit 
explicit results here and proceed with temperatures $T{\textstyle
 {\lower 2pt \hbox{$<$} \atop \raise 1pt \hbox{$ \sim$} }}T_c$ where for
coordinates close to the barrier top a Gaussian approximation -- then
around the newly emerging paths with amplitudes $\pm q_1$ -- is
again appropriate. As discussed above all real time paths with
$q_f<q_1$ have now the 
same oscillation period $t_p(q_1)\approx \ln(q_a/q_1)/\omega_b$. 
One observes that even though they are influenced by the
anharmonicity of the potential via the Euclidian amplitude $q_1$, their
time evolution for $T{\textstyle
 {\lower 2pt \hbox{$<$} \atop \raise 1pt \hbox{$ \sim$} }}T_c$ is
still dominated by parabolic properties. 
 Then,  a somewhat lengthy
algebra leads to the quasi-stationary density matrix 
\begin{equation}
\rho_{\rm fl}(q_f,-q_f)=\frac{1}{2} \rho_\beta(q_f,-q_f)
\left[g_{\rm fl}^{(+)}(q_f,-q_f)+
g_{\rm fl}^{(-)}(q_f,-q_f)\right]\label{nondb3}
\end{equation}
where 
$g_{\rm fl}^{(\pm)}$ describe the contributions from each of the two
oscillating Euclidian paths. Note that due to symmetry  in
$\rho_\beta(q_f,-q_f)$, these
contributions  are identical and just lead to
a factor of 2. In the temperature domain studied here,
$q_1$  can be gained analytically from
Eq.~(\ref{db4a}) as 
\begin{equation}
q_1=\frac{2 q_a}{\sqrt{3}}\,
\left(1-\frac{\pi^2}{\omega_b^2\hbar^2\beta^2}\right)^{1/2}.\label{nondb3a}
\end{equation}
Accordingly, one finds with some algebra for the thermal distribution
\begin{equation}
\rho_\beta(0,0)=\frac{1}{\sqrt{2\pi\delta_b^2|\sin(\omega_b\hbar\beta)|}}\,
\exp\left[\frac{\omega_b \hbar\beta
q_a^2}{12\delta_b^2}\left(1-\frac{\pi^2}{\omega_b^2 \hbar^2 
\beta^2}\right)^2\right].\label{nondb5} 
\end{equation}
The form factor now has two contributions of the form
(\ref{gt18}) and  for  $\kappa_1=\partial\ln(q_1)/\partial
(\omega_b\hbar\beta)\gg  
1$ the corresponding integration boundaries read
\begin{equation}
u^{(\pm)}(q_f) =  \frac{\pm q_1 +i q_f}{
4\delta_b\sqrt{\kappa_1}},\ \ 
u^{\prime\,(\pm)}(q_f,z)=\kappa_1 \left[u^{(\pm)}(q_f)-z\right].\label{nondb4}
\end{equation}
Here, we employed that near $T_c$ the derivative $\Sigma_{ss}$
is dominated by $\partial^2\bar{S}/\partial q_i^2$
which is proportional to $\partial E_1/\partial
(\omega_b\hbar\beta)\propto \kappa_1$. 
This way, using 
 the normalization (\ref{nondb1}), the result for the rate is
\begin{equation}
\Gamma=\frac{\omega_b}{2\pi}
\frac{\sinh(\omega_a\hbar\beta/2)}{\sqrt{2|\sin(\omega_b\hbar\beta)|}}
\sqrt{\omega_b\hbar\beta-\frac{\pi^2}{\omega_b\hbar\beta}}\ \  {\rm
e}^{-\beta V_b}.\label{nondb6}
\end{equation}
This expression is valid for temperatures $T<T_c$ where still
$\kappa_1\gg 1$, a region  which can be estimated as
$T$ somewhat larger than $T_c/2$.
There are two interesting observations to mention: first, the exponentially
large term in the thermal distribution (\ref{nondb5})  -- a
consequence of the 
new Euclidian paths -- is exactly canceled by a corresponding term
which arises from the derivative of the form factor. This way, the
rate is still dominated by the characteristic ``Arrhenius
factor''. Second, in the limit $T\to T_c$ the above-$T_c$ formula
(\ref{nondb2}) and  the below-$T_c$ result (\ref{nondb6}) both
approach $\Gamma_c=(\omega_b/2\pi) \sinh(\omega_a\hbar\beta/2)
\exp(-\beta V_b)$, however, the derivatives $\partial\Gamma/\partial
T$ are different. This discontinuity in the slope of the temperature
dependent rate is removed by the full semiclassical theory \cite{dyn2}
which takes the non-Gaussian fluctuations near $T_c$ into account and
leads to a smooth changeover between the rate formulas (\ref{nondb2}) and
(\ref{nondb6}).

With further decreasing temperature the amplitude $q_1$ tends to
saturate at $q_a$ so that $\kappa_1\to 0$
 and the above rate expression
is no longer 
applicable. Furthermore, the plateau region shrinks and
eventually vanishes so that the assumption of a quasistationary flux
state becomes inadequate even in the sense of a transition state
theory limit of a weak damping theory. 
We note that in case of finite damping a
meaningful rate can be found for much lower temperatures, then
describing  incoherent quantum tunneling. To investigate the time
dependence of $\rho(q_f,-q_f,t)$ with no damping in the limit
of deep tunneling, we consider the case $T=0$ in the next section.

\subsection{Nonequilibrium dynamics for zero temperature}\label{nondb0}

Any Gaussian semiclassics  to the real time propagator is expected to
break down 
for very low temperatures and  very long times where quantum tunneling
comes into play. In 
fact, as yet  no satisfactory semiclassical procedure was found to
describe deep tunneling in the time domain. The crucial question
thereby is: how can  classical trajectories that  either  
oscillate in one of the 
potential wells, here with energy $E< 0$, or move over the barrier,
here with $E>0$,   produce  exponentially small  contributions to
the semiclassical propagator which originate from quantum states
 {\it connecting} the two wells {\it under} the barrier. Here, we
present a 
mechanism that is based only upon the complex plane mechanics
discussed above and avoids any additional {\it ad hoc} insertion of 
``barrier paths''. 
Since  the complex plane dynamics
behaves as the usual classical real time mechanics, paths with $E<0$
never cross the barrier. However, a full semiclassical treatment needs
to account for the dominant fluctuations about the semiclassical
paths. Now,
 for $T<T_c$ 
there is a whole family of loop-like orbits in the complex plane; all
with the same energy, period, and action increase after one period
differing from each other only by their
respective phases, i.e.\ by their crossing points $q_f\leq q_1$ with
the real axis. 
It turns out that  each time these orbits pass their
end-coordinate $q_f$ there are other trajectories of this family
arbitrarily close in phase space (see fig.~\ref{doublephase}). The
role of quantum 
mechanics then is to induce transitions between 
these orbits via small fluctuations. For sufficiently long times a
path starting at a certain  
$q_f$ near the barrier top  may successively slip down to an orbit
 with another phase $q_f'$,  eventually reach the stable regions around
$\pm q_a$, and fluctuate in the long time limit between these
regions. That this scenario  
actually describes the  low temperature coherent tunneling dynamics
has been discussed briefly in \cite{epl} and will be described in some
detail in the sequel.

For $T=0$ the amplitude of the Euclidian time paths is
$q_1=q_a$. Thus, all stationary paths $q_s(q_f,t)$, $q_s'(q_f,t)$ have energy 
$E=V(q_a)$ and period $t_a\equiv t_p(q_a)=2 \pi/\omega_a$. Then, the
Euclidian action $\bar{S}(q_i,q_i')$ suppresses energy fluctuations
around $E=V(q_a)$ exponentially, so that classical paths running in
time $t$ from $q_i\neq q_s$ and $q_i'\neq q_s'$ to $q_f$ and $-q_f$,
respectively, i.e.\ with $E\neq V(q_a)$,  are negligible.
Further, studying the
short action, $W(q,q')=\int_q^{q'} dq\, p\equiv S(q,q')+E t$, one finds
according to Eq.~(\ref{db7a}) that after each period
$W(q_f,q_f)=W(q_a,q_a)=0$. This result combined  with the fluctuation
prefactor 
[see Eq.~(\ref{pre})]  gives for the
Gaussian propagator after multiple round trips and for coordinates $q_f<q_a$
\begin{equation}
|G_{n t_a}(q_f, q_f)|^2\propto \frac{1}{n t_a (q_a^2-q_f^2)},\ \ \ 
n=1,2,3,\ldots.\label{to1} 
\end{equation}
Hence, the probability to return to the starting point decreases as
the  number of periods increases.  In contrast, in the vicinity of the
wells the 
Gaussian propagator coincides with the harmonic propagator. To be more 
precise, due to caustics in the semiclassics of this simple propagator
at all $n 
t_a/2$,  an 
extended semiclassical analysis must be invoked leading to an Airy
function; details  
of the procedure are well-known \cite{schulman} and of no interest
here. The important point 
is that in the barrier region the simple semiclassical return
probability decays to zero for large times while in the well regions it
remains constant.  Thus, we conclude that
the dominant quantum fluctuations neglected in the
Gaussian approximation to the real time propagators are those that
connect stationary paths with the same energy but
 different phases, i.e.\ initial coordinates $q_f$. Effectively,
these relevant fluctuations shift $q$ slightly away from the classical
path $q_s(q_f,t)$ to reach another stationary path $q_s(q_f',t)$
(cf.~fig~\ref{fluczero}). The 
corresponding change in action after a period and for small deviations
is simply
\begin{equation}
W(q_f',q_f)\approx p_s(q_f,0) (q_f'-q_f).\label{to2}
\end{equation}
This repeats at subsequent oscillations. Hence, a ``fluctuation path''
can be characterized by its sequence of crossing points with the real
axis after each round trip, e.g.\ by $q^{(k)}, k=1,\ldots,n$ for $t=n
t_a$ where $q^{(1)}=q_f$. 
Accordingly,  a
fluctuation path is not a classical path, i.e.\ it does not
fulfill Newton's equation of motion, but  can be seen as
 almost classical since it stays always in the close vicinity of a
classical path. 
In the sequel we first explain the general structure of the
extended semiclassical approximation and later on turn to details
of the calculation.

As an example let us consider a fluctuation path starting at
$q_f$ that spirals around $q_a$ 
while the  crossing point $q^{(n)}$ with the real
axis diffuses
close to $q_a$ and returns to $q_f$ in $t\gg t_a$ (see
fig.~\ref{diffusion}a). According to 
(\ref{to2}) on the way to $q_a$ a particular
 path gathers an additional action $W^+(q_a,q_f)$ which is imaginary due
to imaginary $p_s(q,0)$ [see Eq.~(\ref{gt10})] where
\begin{equation}
|W^+(q_a,q_f)|=\int_{q_f}^{q_a} dq\, \left\{2 M
[V(q)-V(q_a)]\right\}^{1/2}\label{to3}
\end{equation}
and the $+\,$ [$-$] sign stands for clockwise
[anti-clockwise] rotation 
of the path in the  complex plane.
As long as the crossing point $q_f$ does not diffuse
 the imaginary time path
connecting the endpoints of the two real time paths coincides at
$t=n t_a$ with the imaginary time orbit
connecting $-q_f$ with $q_f$ at $t=0$.  Taking into account the phase
fluctuations, however, 
forces the endpoint of the imaginary time path  to move
 with  the endpoint $q^{(n)}$ of the ``real time path'' 
 also towards $q_a$. The mapped imaginary time path after $t\gg t_a$
therefore runs from $-q_f$ to $q_a$.
According to
(\ref{gt10}) the additional amount of Euclidian action required for
this deformation of the imaginary time path exactly
counterbalances $W(q_a,q_f)$ so that the total action $\Sigma$ remains
constant which reflects the stationarity of $\Sigma$ along
stationary paths. 
From close to $q_a$ the fluctuation path spirals back to 
$q_f$. However, since $q_a$ is a branching point of the momentum there
are two channels: the real time fluctuation path can maintain 
the direction of rotation or pass the
turning point (TP) 
 $q_a$, thus  changing the sense of rotation (cf.~fig.~\ref{diffusion}a). In
the former case on the way back from $q_a$ to $q_f$ the 
fluctuation path  crosses  the real axis with the 
same direction of momentum as on the way  to $q_a$, so that due to
$W^+(q,q')=-W^+(q',q)$ the path
looses the action $W^+(q_a,q_f)$ again and returns to $q_f$ with
$W(q_f,q_f)=0$. 
In the latter case, momenta on the way back  have
opposite direction to those on the way forth so that  the path arrives
at $q_f$ with action $W(q_f,q_f)=W^+(q_a,q_f)+W^-(q_f,q_a)=2
W^+(q_a,q_f)$ and momentum $-p_s(q_f,0)$. 

Moreover, a fluctuation path starting at $q_f>0$ can
either move along the real axis to the right to reach $q_a$ or
move  to the left to arrive at $-q_a$. In the latter case, the crossing
point $q^{(n)}$ diffuses across the barrier top so that the path
initially spiraling around $q_a$ finally orbits around $-q_a$ with
opposite sense of rotation. Accordingly, due to
$W^+(q_a,0)=-W^-(-q_a,0)$ the real time action factor $\exp[i W(\pm
q_a,0)/\hbar]$  grows or
decreases exponentially for diffusion  to the right or to the left,
respectively. In any case,    
near $\pm q_a$ the
semiclassical propagator has to match onto the  propagators in the harmonic
wells. For the two lowest lying eigenstates which are relevant here,
 the matching procedure was discussed in detail by Coleman
\cite{coleman}.
 Correspondingly,
 these two states determine the  relevant propagator
 in its spectral representation. Then,  
it turns out  that a TP may only occur if
$iW(q_f,\pm q_a)<0$. This has profound consequences on the
extended semiclassical approximation: (i) A relevant fluctuation path
from $q_f$ to $q_f$  
must reach $\pm q_a$ 
rotating clockwise  to have a TP. Then $W^+(q_a,q_f)+W^-(q_f,q_a)=2i
|W(q_a,q_f)|$ and the corresponding
contribution to the propagator has  an exponentially small
factor $\exp[-2 |W(q_a,q_f)|]$. (ii) A fluctuation path with more than
one TP has to
alternatively visit TPs at $\pm q_a$  thereby changing its sense of
rotation repeatedly. In passing from one TP to the next the path
gathers the action $W^-(0,-q_a)+W^+(q_a,0)=W^-(0,q_a)+W^+(-q_a,0)=
i|W(q_a,-q_a)|\equiv i\bar{W}_a$ which coincides with the instanton
action in $\Delta$
introduced in Eq.~(\ref{db6b}).  
Hence,  according to (i) contributions from
fluctuation paths 
with TPs do not play any role for short times. For longer times,
however, they become increasingly important, particularly, since a
fluctuation path 
may spend an arbitrary period of time at the TPs $\pm q_a$ where
$V'(q_a)=0$ before leaving them. The detailed analysis shows (see
below) that for $t\gg t_a$ the phase space
of equivalent fluctuations with one TP is therefore $\propto t$ which
compensates for the exponentially small action factor. Moreover, at each
TP a path gathers an additional Maslov index $\nu\to \nu+1$. Then,
according to (ii) the full density matrix is given by a sum over $\nu,
\nu'$ taking into account the proper order of TPs, i.e.
\begin{equation}
\rho(q_f,-q_f,t)=\sum_{\nu,\nu'\geq 0}\,
\rho_{\nu,\nu'}(q_f,-q_f,t), \label{to4}
\end{equation}
where $\rho_{\nu,\nu'}$ denotes the contribution from relevant
fluctuation paths with $\nu$ TPs in the forward and $\nu'$ TPs in the
backward propagator. Note that $\nu,\nu'$ only label the TPs of the
two real time paths. For each $\nu,\nu'$ one has also to sum over all
imaginary time paths connecting the endpoints of the real time orbits.

To evaluate the sum in Eq.~(\ref{to4}) we start by analyzing the term
with $\nu=\nu'=0$. As discussed above the diffusion of the real time
orbits is then irrelevant and the imaginary time path has to run form
$-q_f$ to $q_f$.
To lowest order in $\Delta$ [see
Eq.~(\ref{db6})]  we have the two imaginary time
paths $\bar{q}_{\pm}(q_f,\sigma)$ emerging at $T_c$ that
connect $-q_f$ with $q_f$ via TPs at $\pm q_a$,
respectively. At $T=0$ other solutions of the imaginary time dynamics
with $r>1\,$ [cf.~Eq.~(\ref{db4a})] just 
contain additional 
 intermediate 
instantons, i.e. imaginary time trajectories connecting $q_a$ with
$-q_a$ or vice versa. For  an equilibrium initial preparation (small)
 fluctuations about the stationary paths towards $\pm q_a$, respectively,
 give identical contributions and  
we thus recover $\rho_\beta(q_f,-q_f)/Z$ [Eq.~(\ref{db6})] in
  the limit $T\to 
0$ with the partition
function
\begin{equation}
Z= 2 \exp[-\beta
V(q_a)-\beta\hbar\omega_a/2]\,\cosh(\hbar\beta\Delta/2).\label{to5}
\end{equation}
For the nonequilibrium preparation, however, only fluctuations towards 
$-q_a$ contribute (fig.~\ref{diffusion}b). This
way, one obtains for 
coordinates $q_f$ near the barrier top 
\begin{eqnarray}
\rho_{0,0}(q_f,-q_f,t)&=& \frac{1}{2} \lim_{\beta\to \infty} \frac{1}{Z}
\rho_\beta(q_f,-q_f)\nonumber\\
&=& \frac{4}{\sqrt{2\pi} \delta_a}\,
\exp[-\bar{W}_a/\hbar]\, \cosh\left(q_f
q_a/4\delta_a^2\right)^2\equiv \frac{1}{2}\, \psi_0(q_f)^2\label{to6}
\end{eqnarray}
where we used  the short action of an instanton
$\bar{W}_a=\bar{W}(q_a,-q_a)$ with $\bar{W}_a=|W(q_a,-q_a)|$ and
$\psi_0(q_f)$ denotes the semiclassical ground state 
wave function in the double well.

The next order  real time paths are those with one TP, i.e.\ $\nu=1,
\nu'=0$ and $\nu=0, \nu'=1$ in Eq.~(\ref{to4}). Thereby,
the real time path $q_s(t)$ makes an excursion from
$q_f$ via a TP at $q_a$ to $-q_a$ in case where at $t=0$ the endpoints
 $-q_f$ and  $q_f$ are connected by an imaginary
 time path $\bar{q}_-$, while it diffuses
 from $q_f$ via a TP at $-q_a$ to $q_a$ 
in case of $\bar{q}_+$. Accordingly, 
 one observes that for an equilibrium initial
preparation   all contributions cancel, e.g.\ the contribution
corresponding to $\bar{q}_-$ with $\nu=1, \nu'=0$ cancels that
corresponding to $\bar{q}_+$ with $\nu=0,\nu'=1$. In fact, it can be
shown in the same way that  for an equilibrium
initial state {\it all} terms in the sum (\ref{to4}) with $\nu,\nu'>0$
vanish. However, for the initial 
preparation (\ref{gt3}) a finite result
follows due to the projection onto the left side of the complex
plane. Hence, both real time orbits have to end near $-q_a$ whereby
one trajectory has a TP at $q_a$. 
 According to the above discussion we gain the following action factors: For
$\nu=1, \nu'=0$ one
has $\exp[-3|W(q_a,0)|/\hbar+q_a q_f/(4\delta_a^2)]$ from the
forward  and $\exp[-|W(q_a,0)|/\hbar+q_a q_f/(4\delta_a^2)]$ from
the backward propagator (cf.~fig.~\ref{diffusion}c) while  $\nu=0,
\nu'=1$ gives 
$\exp[-|W(q_a,0)|/\hbar-q_a q_f/(4\delta_a^2)]$ and 
$\exp[-3|W(q_a,0)|/\hbar-q_a q_f/(4\delta_a^2)]\, $
(cf.~fig.~\ref{diffusion}d), respectively. After 
expanding the integrand in (\ref{gt1}) around $-q_a$ up to second
order, the
ordinary integrations over the initial coordiates are seen to be
restricted to the  
harmonic range around $-q_a$ so that the $\theta$ functions can be put
to 1. Then, the integrals effectively describe the stationary real
time motion of 
the equilibrium well distribution $\rho_\beta(-q_a,-q_a)$.  Combining
these findings yields 
\begin{eqnarray}
\rho_1(q_f,-q_f,t)&\equiv&
\rho_{1,0}(q_f,-q_f,t)+\rho_{0,1}(q_f,-q_f,t)\nonumber\\
&=& i 8\, \Phi(t)\, \exp[-2
|W(q_a,-q_a)|/\hbar]\ \sinh(q_f q_a/2\delta_a^2)\  \rho_\beta(-q_a,-q_a)/Z.
\label{to7} 
\end{eqnarray}
Here $\rho_\beta(-q_a,-q_a)$
includes a sum over multi-instanton contributions of the imaginary
time paths,  resulting for $T\to 0$ in
$\rho_\beta(-q_a,-q_a)/Z=\sqrt{M\omega_a/4\pi\hbar}$.  
Further,  $\Phi(t)\propto t$ takes into account the phase space
contribution from equivalent fluctuation paths connecting $q_f$ via a TP at 
 $q_a$ with $-q_a$. These paths differ only in their ``sojourn times''
 at the TP $q_a$.
To evaluate $\Phi$ we adopt the method outlined in
\cite{weisshaffner} to which 
we also refer for further details and
write
\begin{equation}
G_t(q_a,-q_a)= \int_0^t du\, G_{t-u}(q_a,0)\, \dot{q}_s(0,u)\,
G_u(0,-q_a).\label{to8}
\end{equation}
Since for $t\gg t_a$   the sojourn time is
 exponentially large, one can actually  sum the intermediate time step $u$  
  over the
entire time interval up to negligible corrections.
To calculate the semiclassical propagators in the integrand of  (\ref{to8})
 one exploits that
a fluctuation path moving from $q=0$ to $q_a$ in time $t\gg t_a$ spends
 almost all time
by orbiting in the vicinity of $q_a$ thereby diffusing along the 
classical
stationary paths  (\ref{db7}) with $q_f{\textstyle
 {\lower 2pt \hbox{$<$} \atop \raise 1pt \hbox{$ \sim$} }} q_a$ towards
 the TP. Hence,
the 
time dependence of the propagators is determined only by the asymptotic 
behavior of the fluctuation paths. Then, as already derived in the previous
 paragraph, the actions turn out to be independent of time up to
exponentially small corrections and their sum gives rise to the action factor
  $\exp[-|W(q_a,-q_a)|/\hbar]$ in $G_t(q_a,-q_a)$. 
 For the prefactors one uses the 
representation (\ref{pre}), and exploits the fact that
 $\dot{q}_s(0,u)\sqrt{A_{t-u}(q_a,0)A_u(0,-q_a)}$  depends for
 $t\gg t_a$  on time and temperature only through 
$\exp[-(\hbar\beta+it)\omega_a/2]$
 while
its dependence on the intermediate time step $u$ is exponentially small. This
 way, since both exponential factors are already accounted for in
 (\ref{to7}), 
we arrive at
\begin{equation}
\Phi(t)=G_t(q_a,-q_a) \exp[|W(q_a,-q_a)|/\hbar+(\hbar\beta+it)\omega_a/2]
\label{to8a}
\end{equation}
which leads to
\begin{equation}
\Phi(t)= t\, \frac{4\omega_a q_a}{\sqrt{2\pi} \delta_a}.\label{to9}
\end{equation}
Combining this result with Eq.~(\ref{to7}) we derive the one-TP contribution to
Eq.~(\ref{to4}) as 
\begin{equation}
\rho_1(q_f,-q_f,t)= i t \Delta^2\, \sinh(q_a
q_f/2\delta_a^2)/(\omega_a q_a)\label{to10}
\end{equation}
where the tunnel splitting is specified in (\ref{db6b}).
Likewise, contributions from real time path with more than one TP can
be calculated where the proper order of TPs must be taken into
account. Eventually, only contributions with $i^{2k-1}\Delta^{2k},
k=1,2,\ldots$ survive and the time dependent density matrix
(\ref{to4}) for coordinates near barrier top reads
\begin{equation}
\rho(q_f,-q_f,t)=\frac{1}{2} \psi_0(q_f,-q_f)^2+ i\Delta\, \sin(\Delta
t)\, \sinh(q_a q_f/2\delta_a^2)/(\omega_a q_a).\label{to11}
\end{equation}
Hence, the initial state (\ref{gt3}) develops an imaginary,
time dependent part  from which
 the  tunneling current (\ref{gt5}) is gained as
\begin{equation}
J(t)=\Delta \sin(\Delta t)\label{to12}
\end{equation}
describing coherent tunneling between the wells.
This shows that a systematic semiclassical analysis of the real time
dynamics of the system  covers also low temperature tunneling.
We note that the tunnel splitting coincides exactly with the result of
the instanton approach [see Eq.~(\ref{db6b})] where $\Delta$ is related
to the action of an 
{\it imaginary} time path. Within the real time description the
``instanton dynamics'' is replaced by the above-mentioned diffusion
along the real axis in the complex coordinate plane. 

Before we
conclude this section let us briefly sketch how  the crossover from
 coherent decay to incoherent tunneling  occurs within the present
formalism as the temperature is raised; details of the calculation will be
presented elsewhere. For finite temperatures $T>0$ the energy of the
stationary  
paths is $|\bar{E}_1|=|V(q_1)|<|V(q_a)|$ so that the TPs of the 
fluctuation paths $q_1<q_a$ are shifted towards the barrier
top. Accordingly, $V'(q_1)\neq 0$ and the corresponding actions in
Eq.~(\ref{to8}) are no longer independent of time. Thus,  the  time
interval
 $t\gg t_a$ may
eventually exceed the region where   the integrand gives a
contribution
 so that
$\Phi(t)\to \Phi_{\rm fl}$ saturates. 
 We note that for a system with dissipation basically
the same mechanism, namely, an effective action depending on time via
damping induced correlations,  may cause incoherent tunneling even at $T=0$. 

\section{Eckart barrier}\label{eckart}

As another instructive example we analyze in the
following the transport across a genuine scattering potential, namely,
the so-called Eckart 
barrier
\begin{equation}
V(q)= \frac{V_0}{\cosh(q/L_0)^2}.\label{eck1}
\end{equation}
Here, $V_0$ is the barrier height and $L_0$ the typical
interaction range. We drop the condition $V(q=0)=0$ in this section so
that energies are shifted by $V_0$. In fact, the real time dynamics in
this potential 
is much simpler as in the double well: particles steadily injected from a
thermal reservoir to the left of the barrier built up  a flux across
the barrier that is stationary for all times after a certain transient 
time has elapsed. Thus, the corresponding quantum dynamics is 
described by a barrier transmission rate for all temperatures.
 In a semiclassical expansion we use
$\hbar/\sqrt{2M L_0^2 V_0}$ as the small parameter which demands high
and broad barriers. 

\subsection{Thermal equilibrium and stationary phase
points}\label{eckdyn}

The solution of Newton's equation of motion for the Eckart barrier in
imaginary time  reads \cite{weiper}
\begin{equation}
\bar{q}_0(q_f,\sigma)= L_0 \, {\rm arsinh}\left\{
\sqrt{\frac{V_0-\bar{E}}{\bar{E}}}
\sin\left[\omega(\bar{E})\sigma
-\phi_f\right]\right\}\label{eck2} 
\end{equation}
where we introduced the energy dependent frequency
\begin{equation}
\omega(\bar{E})=\omega_b\ \sqrt{\frac{\bar{E}}{V_0}}\label{eck3}
\end{equation}
with the barrier frequency $\omega_b=\sqrt{2 V_0/M L_0^2}$. Energy
$\bar{E}$ and phase $\phi_f$ are determined by the boundary
conditions $\bar{q}_0(q_f,0)=-q_f$ and
$\bar{q}_0(q_f,\hbar\beta)=q_f$. Accordingly, employing one of these
conditions to fix $\phi_f$, the energy can be evaluated from
\begin{equation}
\omega(\bar{E})\hbar\beta=r\pi+[1+(-1)^r] \phi_f,\ \ \
\label{eck3b} 
\end{equation}
where for given temperature real solutions exist only for a finite
 number of integers $r\geq 0$. Accordingly, this nonlinear equation
 gives the amplitude of the semiclassical
path. As a function of temperature the solutions (\ref{eck2}) and
(\ref{eck3b}) 
resemble those in the inverted double well potential
 (cf.~Sec.~\ref{doubleequi}). Particularly, for $T<T_c$ again all paths with
$r=1$ have  the same 
energy independent
of $q_f$
\begin{equation}
\bar{E}_1 \equiv V(q_1)=\frac{\pi^2
V_0}{\omega_b^2\hbar^2\beta^2}\label{eck3c} 
\end{equation}
 with  amplitudes $\pm q_1$ and the same
 frequency 
$\omega_1=\omega(\bar{E}_1)$. 
An
important difference to the double well case, however, is that the
 amplitude $q_1$ of the paths 
 here grows 
without any limit as $T\to 0$, i.e.\ $\bar{E}_1\to 0$. Hence, the
equilibrium density 
matrices $\rho_\beta(q_f,-q_f)$ differ qualitatively in the deep
tunneling regime. While for the double well potential near $T=0$
contributions 
from all multi-instanton paths must be summed up [see
 Eq.~(\ref{db6})], here, the 
density matrix is 
dominated by the oscillating paths newly emerging around
$T_c=\omega_b\hbar/k_{\rm B} \pi$ 
for {\it all} $T<T_c$. For further details of the Euclidian
semiclassics we refer to \cite{weiper}. 

Now, for the stationary phase points  we obtain
\begin{equation} q_s(t)=L_0\, {\rm arsinh}\left\{
\sqrt{\frac{V_0-\bar{E}}{\bar{E}}}
\sin\left[\phi_f-i\omega(\bar{E}) t\right]\right\}, \ \ \
q_s'(t)=-q_s[(-1)^{r+1}t]\label{eck4}  
\end{equation}
 which are connected by the Euclidian path
$\bar{q}_s(\sigma)= \bar{q}_0[\sigma+i(-1)^{r+1}t]$.   Starting at
$q_f\,$ [$-q_f$]  the path $q_s(t)\,$ [$q_s'(t)$] describes for
large times an almost free motion parallel to the real axis
in accordance with the   asymptotically vanishing interaction $V(q)\to
0$ as $q\to \pm 
\infty$. The energy of $q_s(t)$ is controlled by temperature
where qualitatively the two ranges $T>T_c$ and $T\leq T_c$ must be
distinguished. In the first case, $\bar{E}$ depends on $q_f$ and for
$q_f> 0$ we find asymptotically, i.e.\ for  $\omega(\bar{E})t\gg 1$,
\begin{equation}
q_s(q_f,t)\approx L_0\,{\rm arsinh}\left[\frac{\sinh(q_f/L_0)}{2
\sin(\omega_b\hbar\beta/2)} {\rm e}^{\omega(\bar{E}) t}\right]-iL_0
\left(\frac{\pi}{2}-\frac{\omega_b\hbar\beta}{2}\right)\label{eck5}  
\end{equation}
so that for $T>T_c$ all
stationary paths are restricted to the strip $i
(L_0/2)[-(\pi-\omega_b\hbar\beta),(\pi-\omega_b\hbar\beta)]\, $
(cf.~fig.~\ref{eckartor}).  In the  
 range $T< T_c$ and for end-coordinates $q_f< q_1$ the
oscillating Euclidian paths determine the energy independent of $q_f$  as
$E=\bar{E}_1\, $ [see
Eq.~(\ref{eck3c})]  with the  frequency
$\omega_1=\omega(\bar{E}_1)$. Hence, in  $T\ll T_c$ {\it all} path
starting in the   
barrier range ($q_f{\textstyle {\lower 2pt \hbox{$<$} \atop \raise 1pt
\hbox{$ \sim$}}}L_0$)  nearly coincide asymptotically (see also
fig.~\ref{eckartor}) 
for $\omega_b t\gg 1$    
\begin{equation}
q_s(q_f,t)\approx q_1+L_0\, \ln\left[\sinh(\omega_1
t)\right]+iL_0\pi/2.\label{eck6} 
\end{equation}
Orbits with $q_f>L_0$ come also close to $i L_0\pi/2$ but on a larger time
scale, e.g.\ for $q_f$ close to $q_1\gg L_0$ on the scale $1/\omega_1$.
Since  Im$\{q_s'\}\to
-iL_0\pi/2$ for large $t$ we conclude
that the real time stationary phase dynamics for $T<T_c$ is restricted to the
strip $[-iL_0\pi/2,iL_0\pi/2]$ in the complex plane (fig.~\ref{eckartor}). To
complete this 
discussion we address the special cases $q_f=q_1$ and $q_f=0$,
respectively. In the former case the motion starts with zero
momentum and takes place along the real axis, asymptotically
 ($\omega_1 t\gg 1$) as
\begin{equation}
q_s(t,q_1)\approx q_1 +L_0\omega_1 t.\label{eck6b}
\end{equation}
 In the latter case the orbit 
 can only be defined as the limiting trajectory of $q_s(t,q_f)$ for
$q_f\to 0$, thus, running along the imaginary axis from $q=0$ to
$q=iL_0\pi/2$ and afterwards parallel to the real axis. 

The described complex plane dynamics depends essentially on the
analytic properties of the potential $V(q)$. Interestingly, in case of
the Eckart barrier one has for complex $q$ the periodicity
\begin{equation}
 V(q)=V(q+i L_0 n \pi),\ \ \ n\ \mbox{integer}.\label{eck7}
\end{equation}
Hence, the complex plane falls into strips $[(2n-1) i L_0 \pi/2,
(2n+1) i L_0\pi/2], n$ integer, parallel to the real axis  each of
which  with identical classical mechanics and corresponding stationary
phase paths $q_s(q_f+iL_0 n\pi,t)$.
As shown above,
for $T<T_c$ the real time dynamics starting from the real axis at
$t=0$ reaches asymptotically the boundaries of the strip $n=0$, while
in the range 
$T>T_c$ it does not. This has crucial impact on the semiclassical 
analysis  for
low temperatures as will be discussed in Sec.~\ref{ecklow}.

\subsection{Stationary flux for high and moderate low
temperatures}\label{eckrate1} 

The systems starts
from an initial state where the thermal equilibrium is restricted to
the left of the barrier, thus extending to $q\to
-\infty$. Accordingly, after a certain transient time  has elapsed the
flux across the barrier remains 
stationary forever. 
Since with increasing time the stationary phase points move away from
the barrier top, for large times fluctuations of the order of $L_0$ or
larger are needed to shift $q_i$ into the region Re$\{q_i\}\leq 0$,
then rendering 
the Gaussian stationary phase approximation insufficient.  Yet, we can use
Eq.~(\ref{gt17}) to gain $J_{\rm fl}$ as long as the flux becomes stationary
on a time scale within which $q_s$ remains smaller than  $L_0$.

In the range $T>T_c$ and for $q_f$ near the barrier top one has
$\bar{E}\approx V_0$ so that $\omega(\bar{E})\approx \omega_b$. Then,
the density matrix $\rho(q_f,-q_f,t)$ tends to stationarity on the
scale $1/\omega_b$ while $q_s(t)$ reaches $L_0$ on the much longer
time scale ln$[V_0/(\bar{E}-V_0)]$ only.  We thus regain within this time
window approximatively the 
parabolic result (\ref{para6})  in the semiclassical limit -- large
$V_0$ and $L_0$ -- where anharmonicities are negligibly
small. Correspondingly, the 
rate reads as 
specified in Eq.~(\ref{para7}) with $\omega_b=\sqrt{2 V_0/M
L_0^2}$.

 For temperatures $T<T_c$ the
transient time range grows according to $1/\omega_1=\hbar\beta/\pi$,
 while the upper bound for the validity of the
Gaussian approximation eventually shrinks to $1/\omega_b$. Hence,
 while one can no longer
use the local barrier  dynamics  for temperatures $T\ll T_c$,  in
a region sufficiently close to $T_c$ a rate calculation along the
lines described in Sec.~\ref{nondb} still makes sense. Accordingly, for
$T{\textstyle {\lower 2pt \hbox{$<$} \atop \raise 1pt \hbox{$ \sim$}
}}T_c$  the 
density matrix is obtained as in  Eq.~(\ref{nondb3}) with the
amplitude $q_1$ derived from
Eq.~(\ref{eck3b}) for $r=1$ as
\begin{equation}
q_1= L_0 \, {\rm arsinh}\left[
\frac{\sqrt{(\omega_b\hbar\beta)^2-\pi^2}}{\pi}\right].\label{eck8} 
\end{equation}
This way, one gets the rate 
\begin{equation}
\Gamma=\frac{\omega_b}{4 \pi^2 Z}\
\frac{\sqrt{(\omega_b\hbar\beta)^2-\pi^2}}{{\rm arsinh}\left[
\sqrt{(\omega_b\hbar\beta/\pi)^2-1}\right]}\ \exp\left(-\beta
V_0\right)\label{eck9} 
\end{equation}
for temperatures below $T_c$ but still
above $T_c/2$.  
For even lower temperatures higher order terms in the expansion around
the stationary phase points $q_s,q_s'$ must be taken into account. In
the following section we show  that for $T$  below $T_c/2$ the rate is 
dominated by quantum tunneling which requires an extended
semiclassical analysis. Thus, a higher order expansion
is needed only in the close vicinity of  $T_c/2$ where the
changeover from the thermal to the quantum rate
occurs.

\subsection{Stationary flux for low temperatures}\label{ecklow}

The breakdown of the Gaussian stationary phase approximation for lower
temperatures  indicates also a breakdown of the simple
semiclassical approximation to the real time propagators for large
times. In fact, one needs to carefully analyze the  quantum
 fluctuations around 
the classical paths to capture tunneling processes. In the sequel we
proceed in the spirit of Sec.~\ref{nondb0} and search for   relevant
phase fluctuations.

We begin by recalling that in the range
 $T\ll T_c$ and for coordinates $q_f<q_1$  the classical mechanics 
in the complex plane takes place in strips $[(2n-1) i L_0 \pi/2,
(2n+1) i L_0\pi/2], n$ integer, parallel to the real axis. One thus has
families of classical paths (cf.~fig.~\ref{eckartor}) all with the same
energy $\bar{E}_1$ that 
start at $t\to -\infty$ to the far right  on the lines
$(2n-1) i L_0\pi/2$, run close together with 
almost vanishing momentum $-M L_0 \omega_1$  towards the barrier top,
pass at $t=0$ the coordinates $q_f+i n L_0 \pi$ and then leave again
to the far right moving close together with momentum $M L_0 \omega_1$ 
 asymptotically
along the lines $(2n+1)iL_0\pi/2$.  
Accordingly, for $T\to 0$ in classical phase space, see
fig.~\ref{eckphase},  orbits with
different phases $\phi_f$, i.e.\ different $q_f$,  but from
the same or from adjacent strips lie 
arbitrarily close  to each other in the asymptotic range where
$|V(q)|\to 0$. The effort of
quantum fluctuations then is to link these paths which reflects
the asymptotically free particle diffusion in the Eckart potential. In simple
 semiclassical approximation one has asymptotically the propagator
\begin{equation}
G_t(q,q')= \left(\frac{M}{2 \pi i \hbar t}\right)^{1/2}\
\exp\left[iM\frac{(q-q')^2}{2\hbar t}\right]\label{ecklow1a}
\end{equation}
so that for fixed $q-q'$ and  large times the transition probability
decreases as
$|G_t(q,q')|^2\propto 1/t$ [cf.~Eq.~(\ref{to1})].
Correspondingly,  two different types
 of  fluctuations can be identified: one type of fluctuations connects paths
 $q_s(q_f+i L_0 n\pi,t)$ and
$q_s(q_f'+i L_0n \pi,t)$  within the same
strip, while  the other type of fluctuations  switches between paths
$q_s(q_f+iL_0 n\pi,t)$ 
and $q_s(q_f'+i L_0 (n+1)\pi,t)$ in adjacent strips.
The first type is already  accounted for in the simple
semiclassical approximation to the real time propagator since these
fluctuations  never leave strip $n=0$ and stay in the close vicinity
to the asymptotic 
$q_s(q_f,t)$.  In contrast,  the second type
is relevant beyond the Gaussian semiclassics since it
 causes large deviations
and allows a path $q_s(q_f,t)$ by subsequently diffusing to
another strip to reach a path $q_s(q_f'+iL_0 n\pi,t)$ with $q_f'$ far
from $q_f$ and  $n$ large. Interestingly, this
second kind of fluctuations does not exist for $T>T_c$ where
asymptotically there is always a gap $i\omega_b\hbar\beta$ between paths in
adjacent strips (see fig.~\ref{eckartor}). 

As an example let us consider a
trajectory $q_s(q_f,t)$ with $q_f$ close to the barrier top for $T\ll
T_c$ and times $t\gg 1/\omega_1\gg 1/\omega_b$.  In this 
limit the orbit runs for $\omega_b t\gg 1$ along the boundary $i
L_0\pi/2$ of the strip 
$n=0$ where
fluctuations of the second class bridge the tiny gap to an orbit
$q_s(q_f'+iL_0\pi,t)$ with a  different $q_f'$ in the strip
$n=1$. This trajectory passes $q_f'+iL_0\pi$, 
 and exploiting the periodicity of $V(q)$ the corresponding change in action
 $W(q_f,q_f')$ is shown to read as in Eq.~(\ref{to2}). 
Obviously, the described fluctuations
always lead from an outgoing to an
ingoing orbit thereby increasing the strip number
 which in turn requires a momentum
fluctuation of order  $2|\dot{q}_s(t)|=2M L_0\omega_1$. Estimating
typical momentum fluctuations by 
$\hbar/L_0$ one rederives from $\hbar/L_0\gg M L_0 \omega_1$ the condition
$T\ll T_c$ so that at low temperatures these fluctuations will indeed
occur. By the same procedure the path $q_s(q_f'+i L_0 \pi,t)$ can 
be linked to a path $q_s(q_f''+i L_0 2\pi,t)$ and so forth.
 Similar as in 
case of the double well potential a ``fluctuation path'' is
characterized by its sequence of crossing points $q^{(k)}+i L_0 k\pi$,
$k=0,1,2,\ldots, n\, $ [$q^{(0)}=q_f$], with the lines $i L_0 k\pi$, i.e.\
the copies of  the real axis in the strips $k$. Accordingly,  for
very large times $t\gg 1/\omega_1$ the point $q^{(n)}+iL_0 n \pi$
moves with increasing $n$ along 
the positive imaginary axis 
while simultaneously $q^{(n)}$ can slide down the 
real axis and away from the barrier top to reach the proximity of $q_1$.
From close to $q_1$ relevant fluctuation
paths traverse $q_1$ as a   turning point (TP) -- $q_1$ is a branch
point for the momentum -- and return via the described scenario to
$q_f$ in the strip $k=0$, however, 
crossing the lines $i L_0 k\pi$ with opposite direction of momentum as
on the way forth. The total
change in action is imaginary and given by $W(q_f,q_f)=
W(q_1,q_f)-W(q_f,q_1)=2 W(q_1,q_f)$, 
$n$ arbitrary but large,  where for $q_f$ close
to the top 
\begin{eqnarray} 
|W(q_1,q_f)| &=&  \int_{q_f}^{q_1} dq \left[ 2M(
V(q)-V(q_1))\right]^{1/2}\nonumber\\
&=&
\frac{\pi V_0}{\omega_b}\left(1-\frac{\pi}{\omega_b\hbar\beta}\right)-\omega_b
M L_0 q_f.\label{ecklow1}  
\end{eqnarray}
In a similar way,  the sequence of $q^{(n)}$ of a fluctuation path starting at
 $q_f$ can move directly towards the barrier top, 
diffuse across the barrier to enter the left halfplane of the complex
plane, and end up in the asymptotic region Re$\{q\}\to -\infty$. Since
in leading order the semiclassical propagator has asymptotically to
match onto the free 
propagator (\ref{ecklow1a}), a TP may only occur if $iW(q_f,\pm
q_1)<0$. Hence, what we discussed in Sec.~\ref{nondb0} [see paragraph
above Eq.~(\ref{to4})] can directly be transfered to the situation here 
and the density matrix can be cast into the same form as in
Eq.~(\ref{to4}). In a notable difference to the double well potential,
however,  the TP $q_1$ here is not an isolated extremum of the
potential meaning that each TP -- for Euclidian and  real time
fluctuation paths as well -- is
{\it not} related to an additional phase factor for equivalent
paths. 

After having elucidated the general structure of the semiclassical
density matrix  we
now turn to the explicit calculation of 
the sum  (\ref{to4}) and begin with the term
$\rho_{0,0}(q_f,-q_f,t)$.
 This matrix element follows by the same arguments as given in
Sec.~\ref{nondb0}. Since the equilibrium density matrix for the Eckart
barrier is dominated
by the oscillating paths newly emerging around $T_c$ for all lower
temperatures, all further contributions from Euclidian trajectories
with $r\neq 1$ in Eq.~(\ref{eck3b}) are negligible. Accordingly, we
find for coordinates around the barrier top
\begin{eqnarray}
\rho_{0,0}(q_f,-q_f)&=&\frac{1}{2}\lim_{T\ll T_c}\frac{1}{Z}
\rho_\beta(q_f,-q_f)\nonumber\\
&=& \frac{1}{Z L_0}\left\{\frac{\pi V_0}{\omega_b^2\hbar^2\beta
[(\omega_b\hbar\beta)^2 
(1-q_f^2/L_0^2)-\pi^2]}\right\}^{1/2}\exp\left[-\frac{\pi
V_0}{\omega_b\hbar}\left(2-\frac{\pi}{\omega_b\hbar\beta}\right)
\right].\label{ecklow2} 
\end{eqnarray} 
Note that in contrast to bounded systems the above density matrix remains
temperature dependent even for $T\ll T_c$. 

To next order real time paths with $\nu=1, \nu'=0$ and $\nu=0,\nu'=1$,
respectively, contribute (cf.~figs.~\ref{diffusion}c,d). For $t\gg
1/\omega_1$ a relevant real time 
fluctuation path with $\nu=1$ starting at $q_f$ moves via a TP at
$q_1+i L_0 n'\pi$, $n'$ large,  to the 
left halfplane 
where it eventually crosses $-q_1+i L_0 n\pi$, $n$ large, to
run along the line $i L_0 n\pi$ and reach $q_i+iL_0 n\pi$ in the far left.
For the segment of the fluctuation path from $q_f$ via a TP to
$-q_1+iL_0 n\pi$ the corresponding action factor is $\exp[-3
|W(q_1,0)|/\hbar+\omega_b M L_0 q_f/\hbar]$. Due to the periodicity
(\ref{eck7})  of
the potential  the segment from $-q_1+iL_0 n\pi$ to
$q_i+iL_0 n\pi$, 
$q_i<-q_1$,  can
just be treated as the corresponding one along the real axis; for very
large times $\omega_1 t\gg 1$, i.e. $|q_i|\gg q_1$ according to
Eq.~(\ref{eck6b}), we then get the action factor $\exp[-iM q^2/2\hbar t]$.
 Hence, the
corresponding relevant real time propagator reads
\begin{equation}
G_t(q_f,q_i)=-i \sqrt{A(q_f, q_i)}\, \exp\left[-\frac{3
|W(q_1,0)|}{\hbar}+\frac{\omega_b M L_0 q_f}{\hbar}+iM
\frac{q_i^2}{2\hbar t}\right].\label{ecklow3}
\end{equation}
Similar, the propagator from $-q_f$ directly to $q_i'+i L_0 n \pi$ is
gained. The crucial point is now that for the integral in Eq.~(\ref{gt1})
there are no longer isolated stationary phase points but rather {\it all}
$q_i, q_i'$   on the line $i L_0 n \pi$ and to the far left of the
barrier top make the
integrand for 
very large times stationary. The ordinary integrals in Eq.~(\ref{gt1})
can thus be seen as sums over stationary phase points $q_i, q_i'$
whereby  their distance is weighted by the asymptotic thermal
distribution, i.e. in leading order the free particle equilibrium
 density matrix
\begin{equation}
\rho_\beta(q, q')=\left(\frac{M}{2\pi \hbar^2\beta}\right)^{1/2}\,
\exp\left[-\frac{M(q-q')^2}{2\hbar^2\beta}\right].\label{ecklow4}
\end{equation}
Accordingly, for
$T\to 0$ one has $\rho_\beta(q_i, q_i')\to \delta(q_i-q_i')$ so that
contributions from $q_i\neq q_i'$ are caused by thermal
 fluctuations at elevated temperatures. 
Further, for large $q_i, q_i'$ and large times the prefactors $A(q_f,q_i)$ and
$A'(-q_f, q_i')$, respectively,  are independent of $q_i, q_i'$, thus
allowing us to carry out the $q_i, q_i'$ integrals over the exponentials
only. Then,   
using $-q_1$ as an upper bound for the asymptotic coordinate range 
it turns out that for $\omega_1 t\gg 1$
the result for the integrals  in leading order is $\pi\hbar t/M$. 
Now, combining all factors we finally obtain the time independent density
\begin{eqnarray}
\rho_1(q_f,-q_f)&=&\lim_{\omega_1 t\gg
1}\rho_{1,0}(q_f,-q_f,t)+\rho_{0,1}(q_f,-q_f,t) \nonumber\\
&=& \frac{i}{Z L_0} \left[\frac{4\pi V_0}{\hbar\omega_b
(\omega_b\hbar\beta)^3}\right]^{1/2}\,
\sinh(2\omega_b M L_0 
q_f/\hbar)\, {\rm e}^{-4 |W(q_1,0)|/\hbar}
\label{ecklow5}
\end{eqnarray}
where $|W(q_1,0)|$ follows from Eq.~(\ref{ecklow1}).
Employing the same procedure, contributions in the sum (\ref{to4}) from
real time paths with more than one TP can be derived, however, 
they contain additional action factors and 
are thus exponentially small compared to $\rho_1$. Hence, the
stationary semiclassical
density matrix 
for low temperatures and very large times is found as
\begin{equation}
\rho_{\rm fl}(q_f,-q_f)=\frac{1}{2 Z}\rho_\beta(q_f,-q_f)
+\rho_1(q_f,-q_f)
\label{eqcklow6} 
\end{equation}
with $\rho_\beta$ as specified in Eq.~(\ref{ecklow2}).
Finally, from Eq.~(\ref{gt5}) we gain the thermal tunneling rate
\begin{equation}
\Gamma=\frac{1}{Z}\,\left[\frac{4\pi V_0\omega_b}{\hbar
(\omega_b\hbar\beta)^3}\right]^{1/2}\,
{\rm e}^{-4 
|W(q_1,0)|/\hbar}. \label{ecklow7}
\end{equation}
This simple formula is applicable as long as $q_1>L_0$, 
 a temperature range which can be estimated by $T$  below $T_c/2$,
 or equivalently
 $\omega_b\hbar\beta> 2\pi$. 
To be precise, there is also a lower bound for the temperature. Namely,
 for $T\to 0$ any semiclassics in the Eckart barrier
breaks down due to the fact that then tunneling
takes place in the low energy range near the base of the barrier where
 the wave length of a
 wave function tends to exceed the width of the barrier.
From the known exact transition probability [see e.g.~\cite{weiper}]
one derives that 
this scenario becomes
relevant for $\omega_b\hbar\beta\gg 2 \pi^4 (V_0/\hbar\omega_b)$ 
corresponding in the semiclassical limit $V_0/\hbar\omega_b\gg 1$ to extremely
 low temperatures. 
In the broad temperature range between these bounds, i.e.\ $2 \pi<
\omega_b\hbar\beta{\textstyle {\lower 2pt \hbox{$<$} \atop \raise 1pt
\hbox{$ \sim$}}} 2\pi^4 (V_0/\hbar\omega_b)$, 
the  above rate expression describes the decay rate with remarkable accuracy
when compared to the exact result  even for
 moderate barrier heights (see fig.~\ref{rate}). Table I presents a
 numerical comparison with results 
 from other approaches. For temperatures above $T_c/2$ the real-time
 semiclassical rate is slightly too small and coincides for $T>T_c$
 with the well-known ``unified'' semiclassical rate formula gained by
 the thermal average over the transmission 
$T(E)=1/\{1+\exp[S(E)/\hbar]\}$, where $ S(E)$ is the
bounce action for $T<T_c/2$.
 The small deviations from the exact rate are due to the fact that
 in the simple version of the theory presented here   anharmonicities of the
 potential are neglected for $T>T_c$ and taken into account only in
 leading order in $T_c>T>T_c/2$. A perturbative expansion in an anharmonicity
 parameter allows for a systematic improvement.
 For the same reason, the
  temperature 
 region around $T_c/2$ is not well described.  In the deep tunneling
 region $T<T_c/2$,  which is notoriously problematic for  real-time
 rate theories,
   our theory performs excellently.
In fact, the low temperature formula (\ref{ecklow7}) turns
 out to be identical to the result derived within  the
 instanton/bounce approach [see e.g.\ \cite{voth2}].
 The bounce is an oscillating Euclidian orbit, periodic in
phase space, which 
connects $q_f=0$ with itself, thus emerging as a solution of
Eq.~(\ref{eck3b}) at $T=T_c/2\,$ ($r=2$). While in imaginary time 
methods the bounce trajectory describes barrier penetration, here, 
 effectively the same tunneling rate arises from {\it fluctuations around real
time paths} the energy of which is fixed by oscillating Euclidian
orbits, closed in 
phase space, emerging at $T_c$ ($r=1$). These latter minimal
action paths solely determine the semiclassical thermal equilibrium for lower
temperatures $T<T_c$, thus establishing within a semiclassical real time
 approach the relation between
 the thermal 
density matrix and the thermal tunneling rate, since  long an open
 question in thermal rate theory. 

\begin{table}
\caption{Transmission factor
$P=\Gamma/\Gamma_{cl}$ for the symmetric Eckart barrier. $\Gamma_{cl}$
is the classical rate and parameters are the same as in
\protect{fig.~\ref{rate}.}} 
\begin{tabular}{cccccc} 
 $\omega_b\hbar\beta$ & $P_{\rm rsemi}^{\rm a}$& $P_{\rm uni}^{\rm b}$
&$P_{\rm QTST}^{\rm c}$&$P_{\rm SQTST}^d$& $P_{ex}^{\rm e}$ \\  
\hline
1.5   & 1.10        & 1.10       & 1.13 & 1.13        & 1.13      \\ 
3     & 1.50        & 1.50       & 1.54 & 1.52        & 1.52     \\ 
5     & 2.98        & 3.84       & 3.18 &  --         & 3.11   \\ 
6     & 3.86        & 8.92       & 5.74 & 2.2         &5.2\\
8     & 21.99        & 17.97      & 29.3 & 11.9   &21.8   \\ 
10    & 136.2      & 132.2      & 248 & 149       &162\\
12    & 1613        & 1606       & 3058 &3006     &1970 \\
16& $6.03\cdot 10^5$ &$6.03\cdot 10^5$ & --&$2.56\cdot 10^6
$&$7.41\cdot 10^5$ \\ 
18    & $1.54\cdot 10^7$ &$1.54\cdot 10^7$ & -- &$9.1\cdot 10^7$
&$1.88\cdot 10^7$  
\end{tabular}
{\footnotesize $^{\rm a} P_{\rm rsemi}$ is the transmission factor as
derived by the 
real-time semiclassical approach presented in this paper}\hfill\break
{\footnotesize $^{\rm b} P_{\rm uni}$ is the transmission factor of the
``unified'' semiclassical 
approach}\hfill\break 
{\footnotesize $^{\rm c} P_{\rm QTST}$ is the transmission factor according
to the simplest version of Pollak's QTST,
from Ref.~\cite{pollak2}}\hfill\break
{\footnotesize $^d P_{\rm SQTST}$ is the  transmission factor
according to  the full semiclassical version of Pollak's QTST, from
Ref.~\cite{pollak2}}\hfill\break 
{\footnotesize $^{\rm e} P_{\rm ex}$  is the exact transmission
factor}\hfill\break 
\end{table}

This represents substantial progress when compared with other attempts.
While Pollak's new quantum transition state theory \cite{pollak} is
 based on a numerically exact evaluation of the thermal flux, it
suffers from a simple 
 semiclassical approximation to the real time
 propagators. In the associated full semiclassical calculation by
Pollak and Eckhardt
\cite{pollak2}   only half of 
the bounce action appears in 
the exponential factor for temperatures below 
 $T_c/2$ and the corresponding
 tunneling rates are  too large (see also Table I). The centroid
method \cite{VCM89} gives the 
 correct action factor, however, its semi-empirical factorization of thermal
 and dynamical contributions leads to a  prefactor which is too small
 for lower temperatures. Finally, from semiclassical real time calculations for
 the Eckart barrier
based on  the simple semiclassical propagator \cite{kesha,grossmann}
tunneling probabilities in the deep tunneling regime cannot be
properly extracted since they depend
 strongly on the initial state.

\section{Conclusions}\label{conclu}

We have developed a unified semiclassical theory that describes the
real time dynamics of quantum statistical systems for all temperatures
including coherent and incoherent processes. 
 Starting from the exact
nonequilibrium dynamics  the approximate density matrix is gained
by employing  semiclassical propagators in real and imaginary time
combined with a stationary phase evaluation. Accordingly,  the 
 relevant classical mechanics takes place in the complex coordinate
plane where the energy of classical real time trajectories is
fixed by the Euclidian orbits determining the equilibrium
distribution. Consequently, real time paths follow by solving an
initial value  rather than a boundary value problem.
While this procedure can be used to study the dynamics for a wide
class of systems and  initial 
conditions, here, we concentrated on the flux across a double 
well potential and an Eckart barrier. 
Then, 
for high to
moderate temperatures the Gaussian approximation suffices to obtain a
stationary flux. In the tunneling domain, however, this approximation
fails and the
complex plane dynamics allows to identify the dominant quantum
fluctuations in the real time propagators. These are zero-mode like
phase fluctuations 
which give rise to a diffusion along
the scaffold of classical orbits. Quantum
tunneling in the real time domain can semiclassically thus be
interpreted
 as a diffusion
process on a certain family of 
classical real time paths. By systematically incorporating the phase
fluctuations
 we managed to derive for
the first time within a real time semiclassical formalism coherent
tunneling dynamics.

 Regarding incoherent decay in the deep tunneling
regime, the theory revealed the connection between  thermal
equilibrium and the tunneling rate upon which thermodynamic rate
formulas are based. In the semiclassical limit flux across the barrier
and equilibrium are linked via
 the intimate relation between
Euclidian and real time paths in the complex plane mechanics. 
However, the theory not
only reproduces the results of  various other semiclassical  thermal
rate theories for higher temperatures, but covers with no further
assumptions also the low temperature regime where so far other real
time methods failed. 

Several questions  could not be analyzed in detail in this article:
There is first the temperature range around $T_c/2$ where  approximately
high and low temperature semiclassics match; we briefly sketched
corresponding improvements. Second, we only
touched in passing  the explicit real time dynamics in
the transient time domain where for incoherent processes the
relaxation to a stationary flux occurs. Third, other initial
preparations, e.g.\ to gain correlation functions, were out of the
scope of this paper. 

Moreover, with the appropriate formalism at hand further extentions
are possible. While the dynamics of dissipative systems was already
studied in the high to moderate temperature range
\cite{dyn1,dyn2}, the low temperature 
tunneling regime is now in principle open for investigations. 
Of course, a crucial point for all further
applications  is to
develop  an appropriate numerical algorithm to 
mimic the ``diffusion'' in the complex plane. We hope to make progress
in this direction in the near future.

\section*{Acknowledgments}

We thank P.\ Pechukas for many interesting discussions.
JA acknowledges a Feodor Lynen fellowship of the Alexander von
Humboldt Foundation. Further support was provided by
the DAAD and the DFG through SFB276.

\begin{figure}
\caption{Loop of stationary imaginary and real time
paths in the complex time plane $z=u+i \sigma$.} 
\label{loop}
\end{figure}

\begin{figure}
\caption{Semilcassical paths (dashed lines) in the complex plane near 
the parabolic barrier top. Shaded area contains relevant intermediate
 coordinates $q_i,
q_i'$ reached by fluctuations along arrows.}
\label{hightc}
\end{figure}

\begin{figure}
\caption{Real time paths in the double well potential
with wells at $\pm q_a\, $ (dots)
for various $q_f$ and $T=0$ (thin lines). The thick line shows a
typical fluctuation connecting orbits with different $q_f$.}
\label{fluczero}
\end{figure}

\begin{figure}
\caption{Phase spaces orbits of complex real time paths in  the double
well potential at $T=0$. In the left picture the real part of the orbits 
is shown for trajectories starting with $q_f>0$ near the barrier top;
the dot indicates the well at $q_a$. In the right picture the corresponding
imaginary parts are  depicted.}
\label{doublephase}
\end{figure}

\begin{figure}
\caption{Diffusion of the crossing point $q^{(n)}$ of a
fluctuation path along the real axis (thick lines) for various cases
discussed in the 
text. Dots indicate the wells at $\pm q_a$ that are branch points for the
 momenta, thin vertical lines the
end-coordinates at $\pm q_f$. Solid lines refer to the forward, dotted
ones to the backward propagator, a crossing of a dot a TP.} 
\label{diffusion}
\end{figure}

\begin{figure}
\caption{Real time paths in the complex plane for the
Eckart barrier. Solid lines show orbits for $T<T_c$, dotted lines
orbits for $T>T_c$.}
\label{eckartor}
\end{figure}

\begin{figure}
\caption{Phase space orbits of complex real time paths in the
Eckart barrier potential at $T\ll T_c$. In the left picture the
 real part of the orbits 
is shown for trajectories starting with $q_f>0$ near the barrier
top. In the right picture the corresponding 
imaginary parts are depicted; the dotted line separates the
strips $n=0$ and $n=1$.}
\label{eckphase}
\end{figure}

\begin{figure}
\caption{Transmission factor $P$ as a function of inverse temperature for an
Eckart barrier with  $\alpha=12$, $\alpha=2\pi V_0/\hbar\omega_b$.
$P$ is defined by
$P=\Gamma/\Gamma_{cl}$ with $\Gamma_{cl}$ the classical rate, i.e.\
the high temperature limit to Eq.~(\ref{para7}). The solid line is the exact
result. In the left picture the dotted line shows the parabolic result
Eq.~(\ref{para7}), 
the dashed line represents  Eq.~(\ref{eck9}), and the arrow indicates
the inverse temperature corresponding to $T_c$. 
 In the right picture the dashed line depicts
 the result Eq.~(\ref{ecklow7}) and the arrow refers to $T_c/2$.}
\label{rate}
\end{figure}

\pagebreak

\setlength{\unitlength}{1cm}

\begin{picture}(0,17)

\put(-2.5,-6){
\makebox(22,22){
               \epsfysize=25cm
               \epsffile{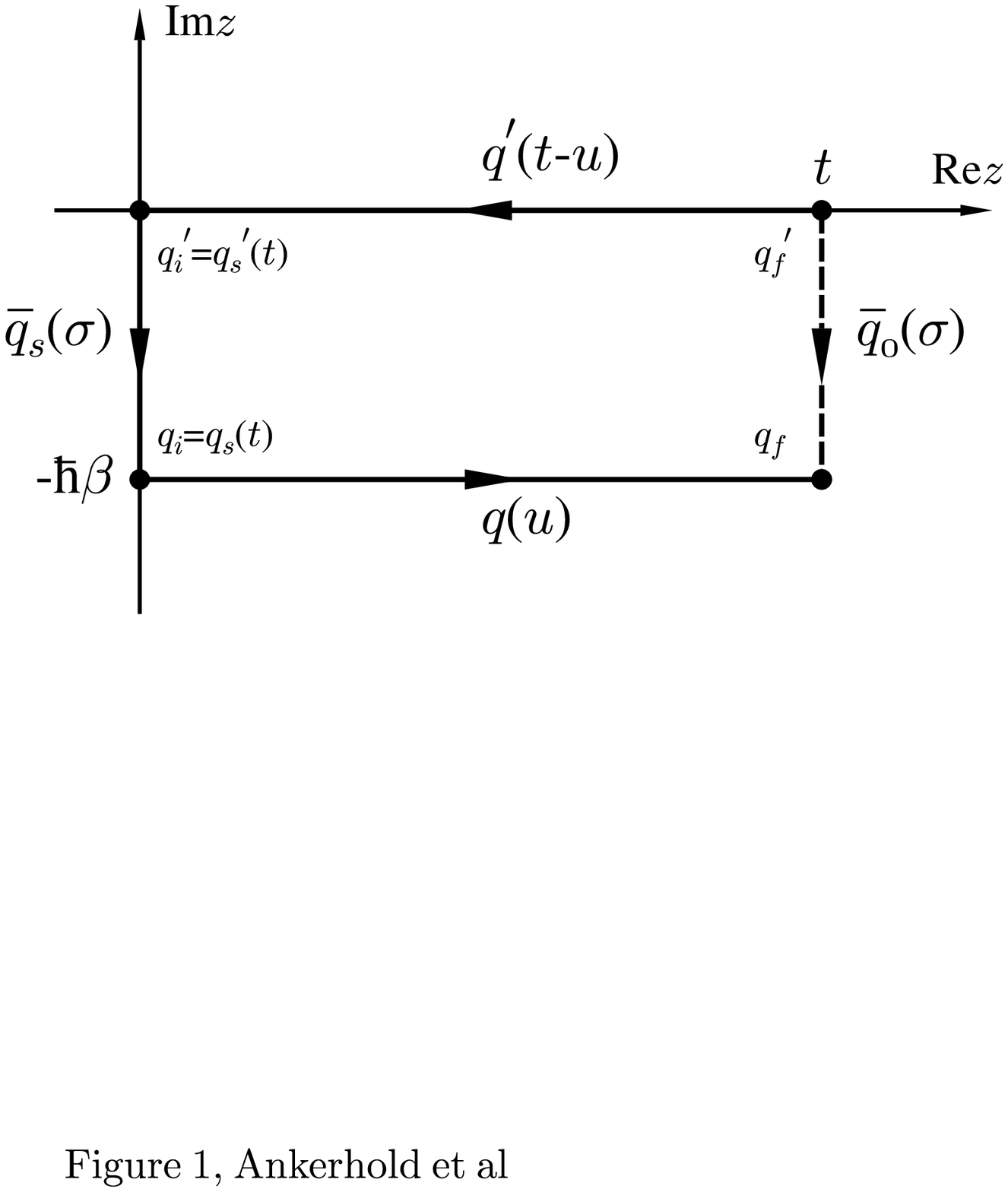
                        }
                          }
          }

\end{picture}

\pagebreak

\begin{picture}(0,17)

\put(-2.5,-6){
\makebox(22,22){
               \epsfysize=25cm
               \epsffile{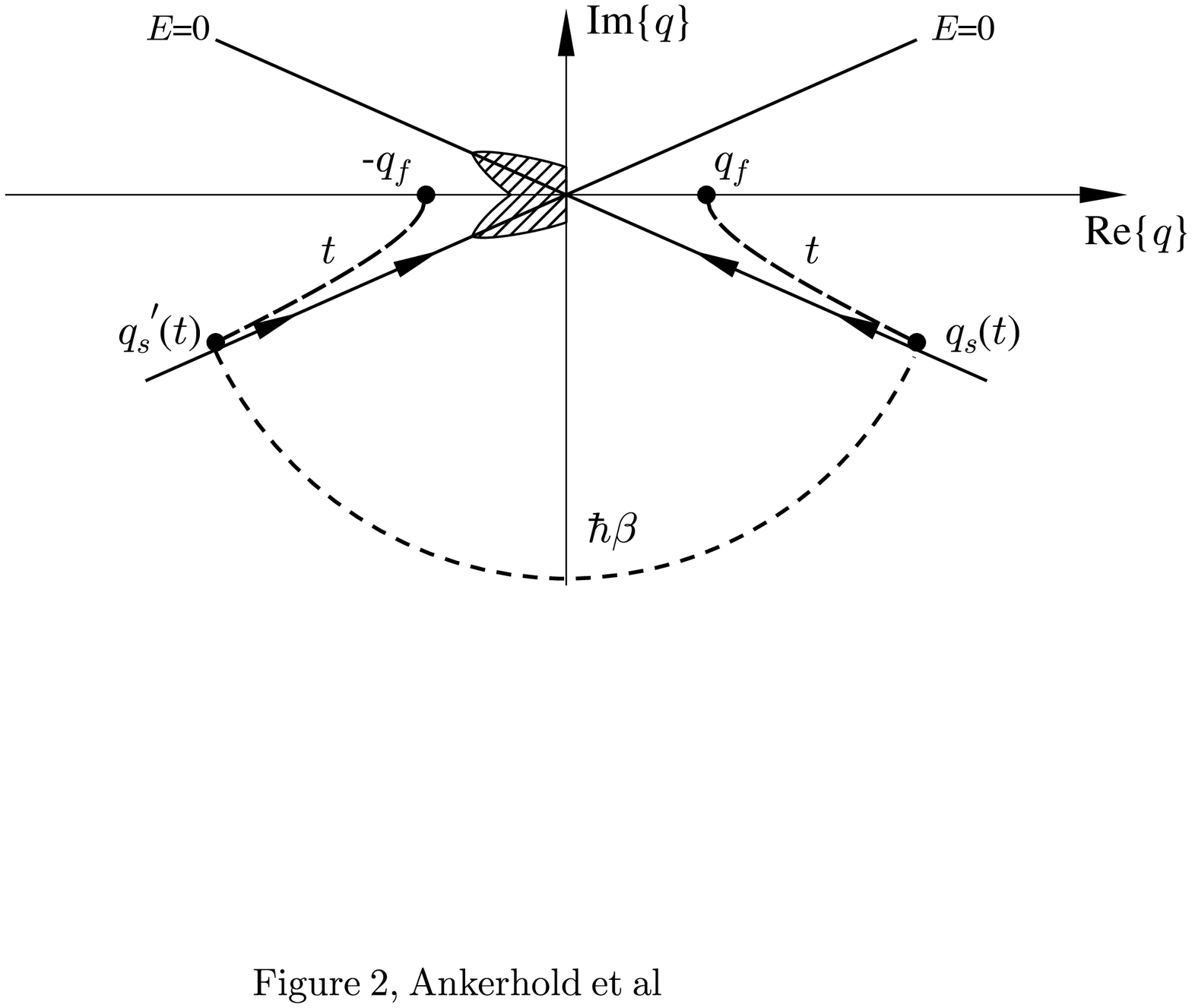
                        }
                          }
          }

\end{picture}

\pagebreak

\begin{picture}(0,17)

\put(-2.5,-9){
\makebox(22,22){
               \epsfysize=18cm
               \epsffile{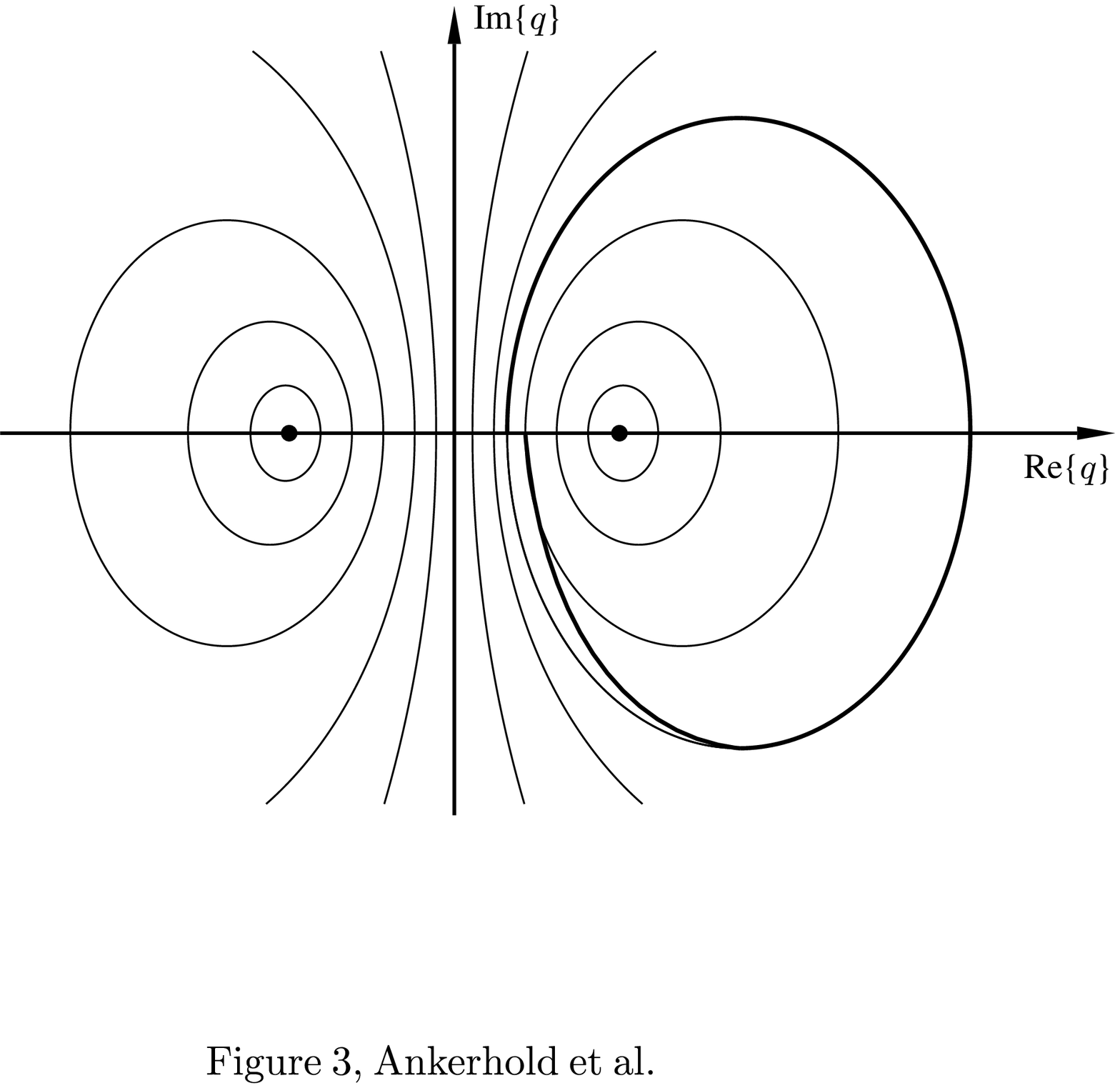
                        }
                          }
          }

\end{picture}

\pagebreak

\begin{picture}(0,17)

\put(-3,-6){
\makebox(22,22){
               \epsfysize=25cm
               \epsffile{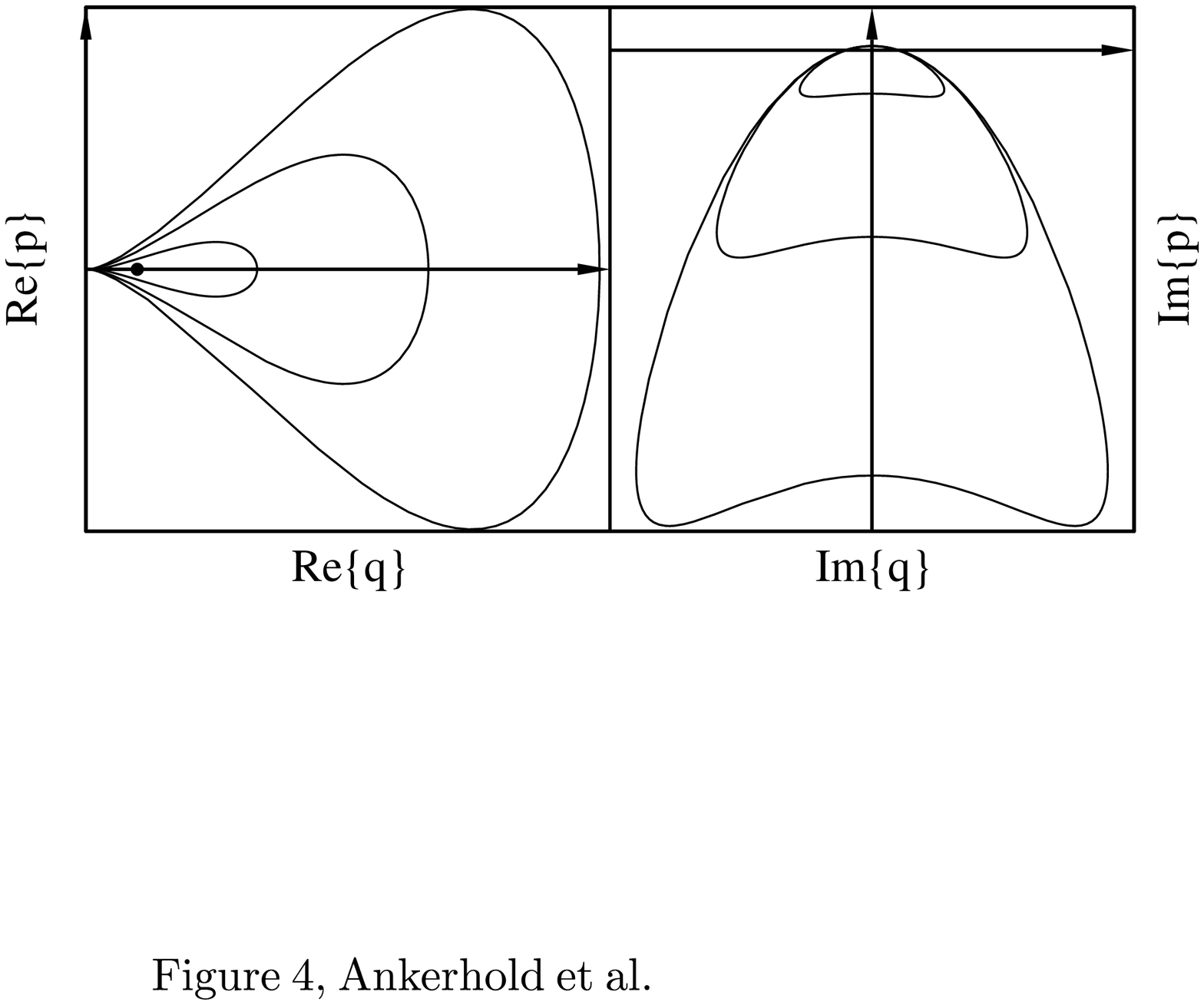
                        }
                          }
          }

\end{picture}

\pagebreak

\begin{picture}(0,17)

\put(-3.5,-6){
\makebox(22,22){
               \epsfysize=25cm
               \epsffile{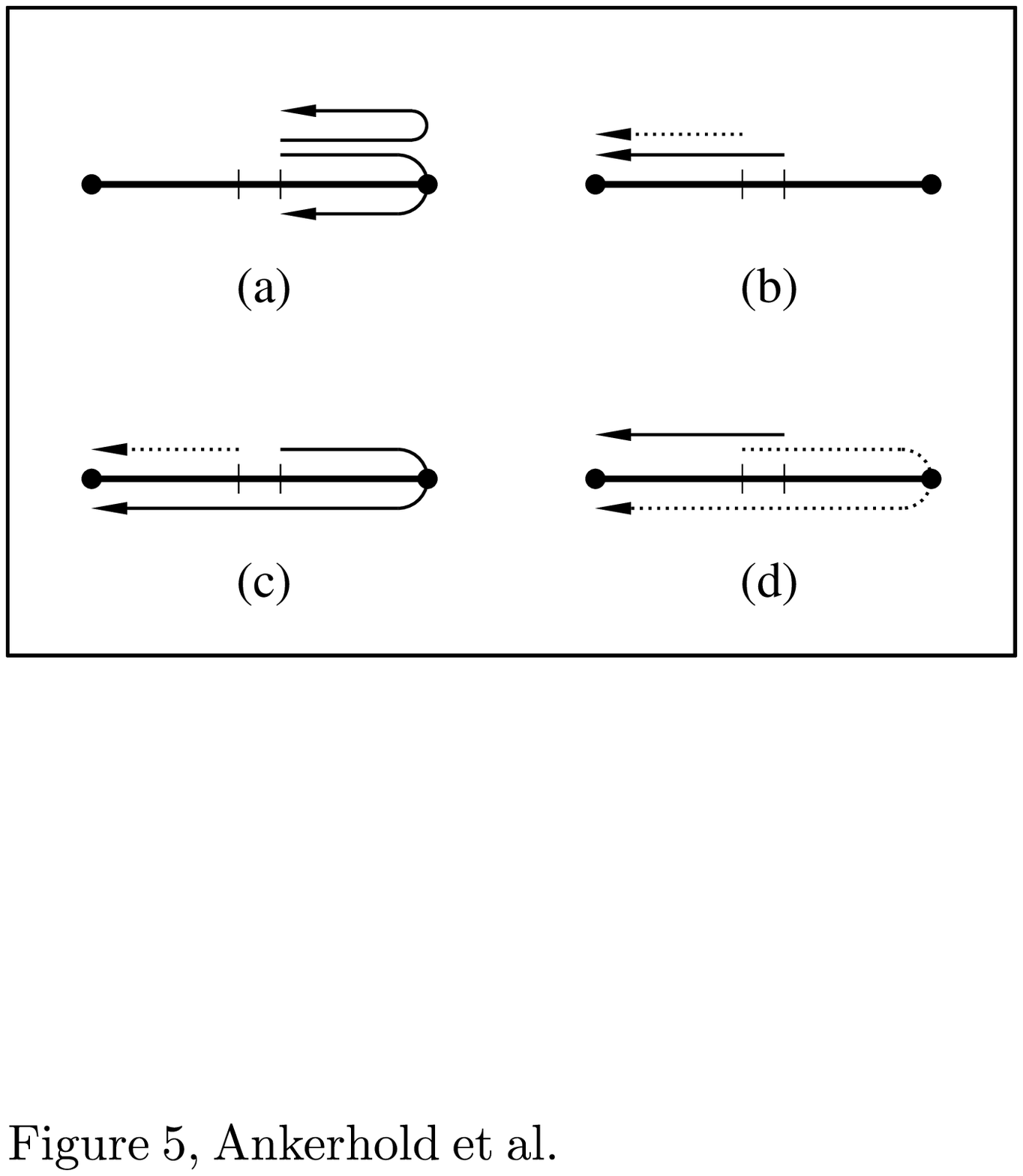
                        }
                          }
          }

\end{picture}

\pagebreak

\begin{picture}(0,17)

\put(-3.5,-6){
\makebox(22,22){
               \epsfysize=25cm
               \epsffile{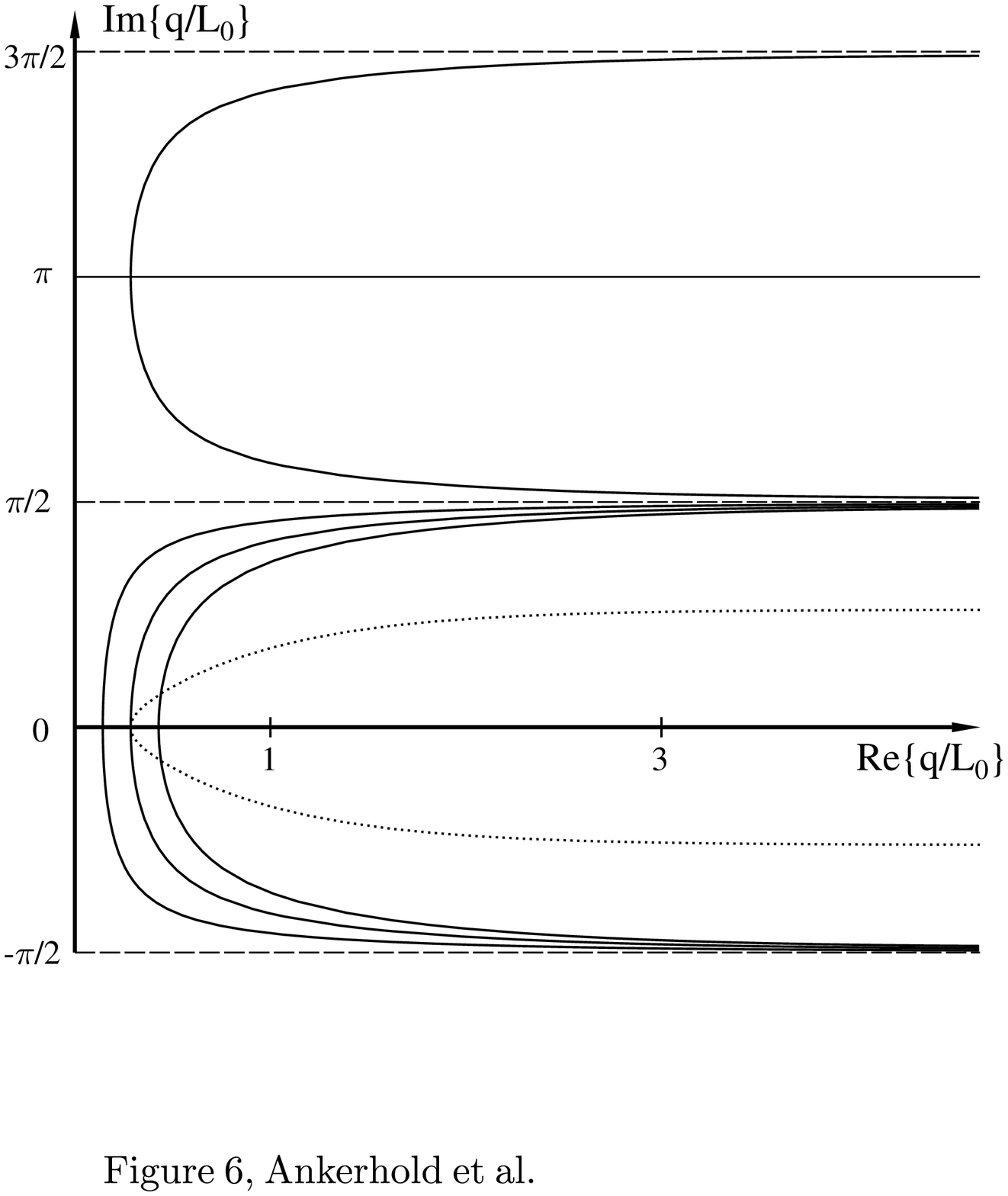
                        }
                          }
          }

\end{picture}

\pagebreak

\begin{picture}(0,17)

\put(-3,-6){
\makebox(22,22){
               \epsfysize=25cm
               \epsffile{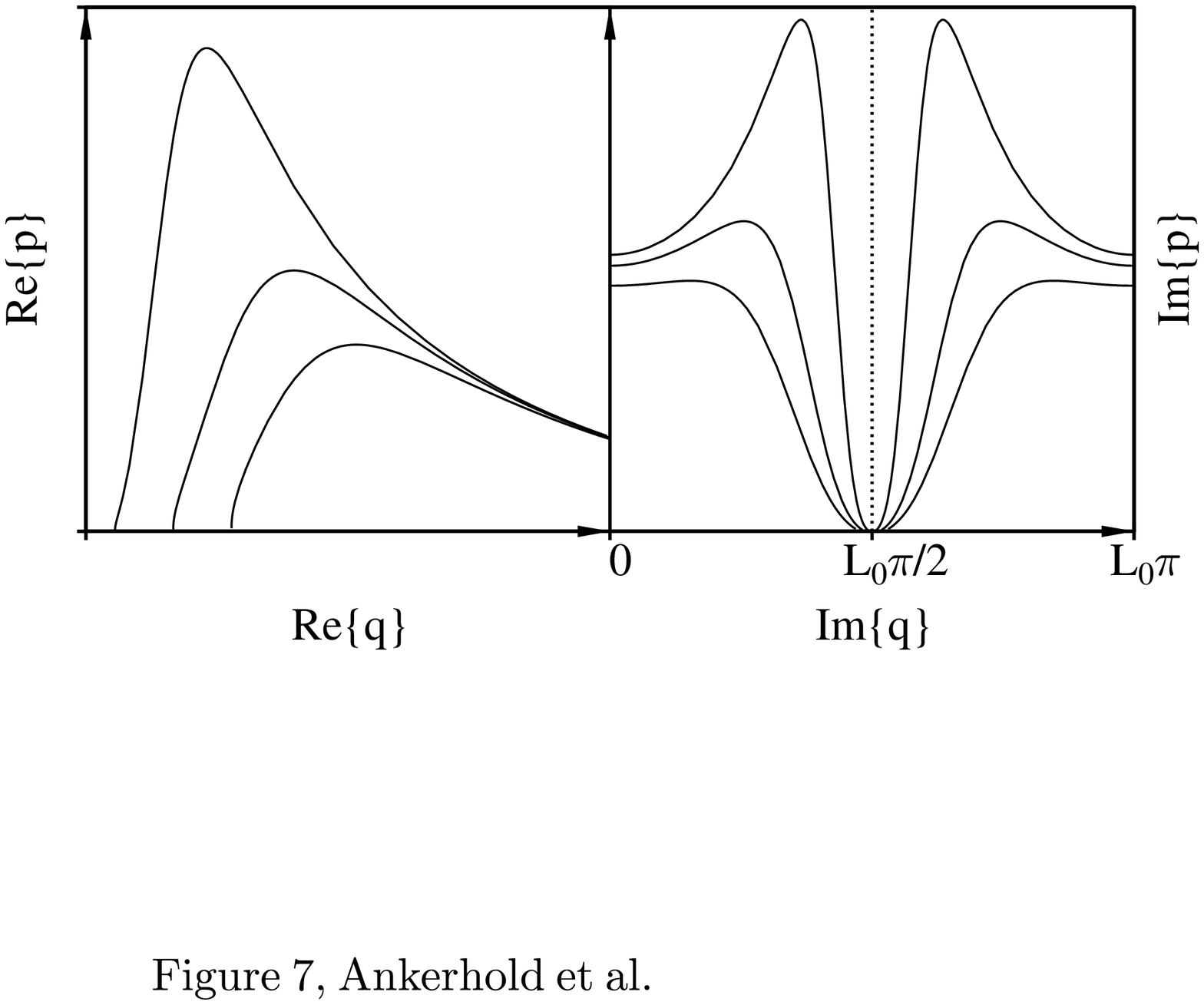
                        }
                          }
          }

\end{picture}
 
\pagebreak

\begin{picture}(0,17)

\put(-2,-6){
\makebox(22,22){
               \epsfysize=25cm
               \epsffile{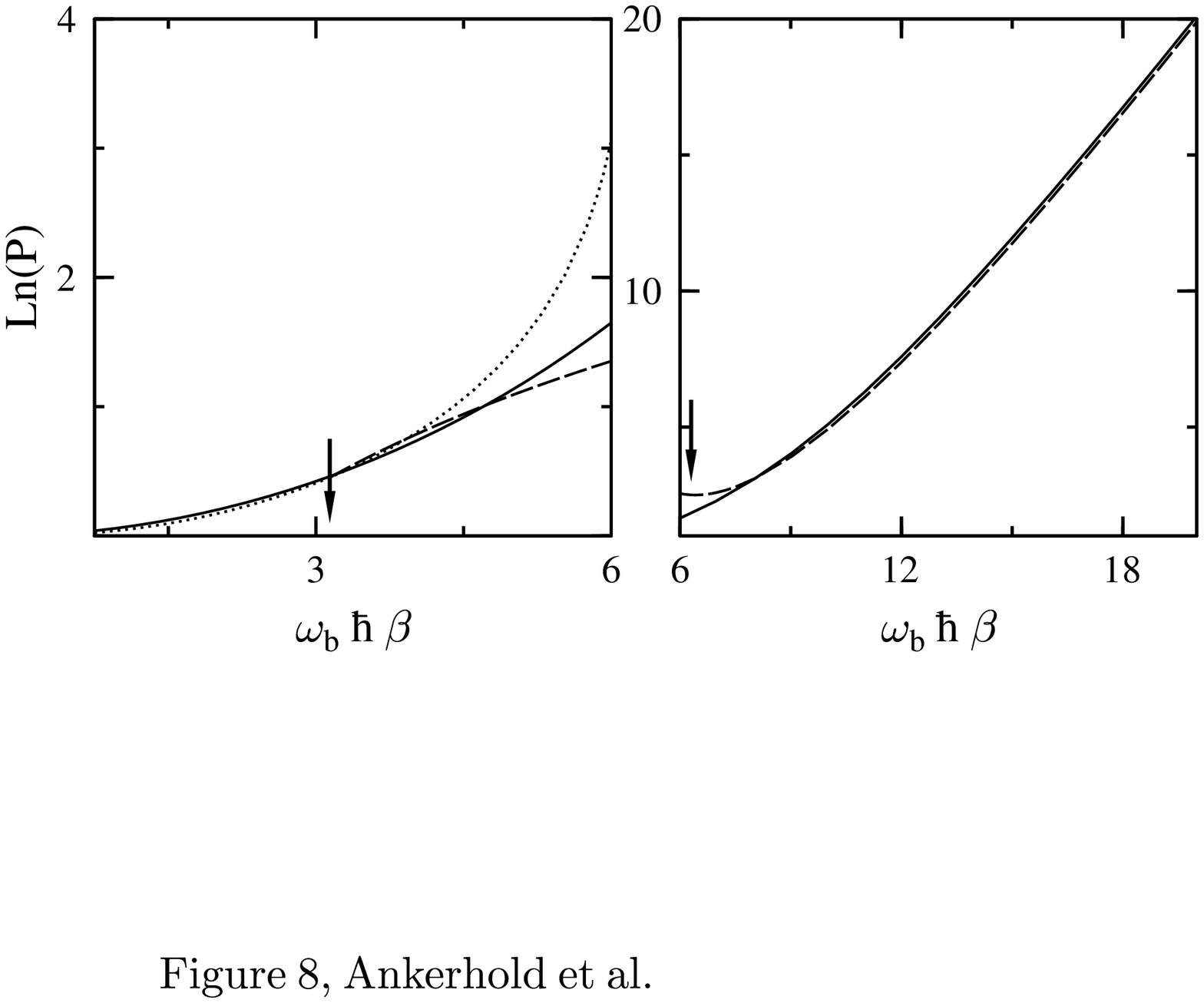
                        }
                          }
          }

\end{picture}

\end{document}